\documentclass[aps,prd,preprintnumbers,groupedaddress,nofootinbib,showpacs,amssymb,eqsecnum,notitlepage]{revtex4}
\usepackage{here}
\usepackage[dvipdfmx]{graphicx}
\usepackage{amsmath,amsthm,amssymb}
\usepackage{bm}
\usepackage{color}

\usepackage{amsfonts}
\usepackage{dcolumn}
\usepackage{hyperref}
\allowdisplaybreaks[1]

\begin{document}
\newcommand{\newc}{\newcommand}

\newcommand{\SN}[1]{{\color{red} #1}}
\newcommand{\ben}{\begin{eqnarray}}
\newcommand{\een}{\end{eqnarray}}
\newc{\be}{\begin{equation}}
\newc{\ee}{\end{equation}}
\newc{\ba}{\begin{eqnarray}}
\newc{\ea}{\end{eqnarray}}
\newc{\bea}{\begin{eqnarray*}}
\newc{\eea}{\end{eqnarray*}}
\newc{\tp}{\dot{\phi}}
\newc{\ttp}{\ddot{\phi}}
\newc{\nrhon}{n_I\rho_{I,n_I}}
\newc{\nrhocn}{n_c\rho_{c,n_c}}
\newc{\drho}{\delta \rho_I}
\newc{\drhoc}{\delta \rho_c}
\newc{\dphi}{\delta\phi}
\newc{\D}{\partial}
\newc{\ie}{{\it i.e.} }
\newc{\eg}{{\it e.g.} }
\newc{\etc}{{\it etc.} }
\newc{\etal}{{\it et al.}}
\newcommand{\nn}{\nonumber}
\newc{\ra}{\rightarrow}
\newc{\lra}{\leftrightarrow}
\newc{\lsim}{\buildrel{<}\over{\sim}}
\newc{\gsim}{\buildrel{>}\over{\sim}}
\newc{\aP}{\alpha_{\rm P}}
\newc{\delj}{\delta j}
\newc{\rhon}{\rho_{m,n}}
\newc{\rhonn}{\rho_{m,nn}}
\newc{\delrho}{\delta \rho_m}
\newc{\pa}{\partial}
\newc{\E}{{\cal E}}
\newc{\rH}{{\rm H}}
\newc{\ty}{\tilde{y}}

\preprint{YITP-20-56, WUCG-20-02}

\title{Suppressed cosmic growth in coupled vector-tensor theories}

\author{
Antonio De Felice$^{1}$, Shintaro Nakamura$^{2}$, and  
Shinji Tsujikawa$^{3}$
}

\vspace{1cm}

\affiliation{
$^1$Center for Gravitational Physics, Yukawa Institute for 
Theoretical Physics, 
Kyoto University, 606-8502, Kyoto, Japan\\
$^2$Department of Physics, Faculty of Science, 
Tokyo University of Science, 1-3, Kagurazaka,
Shinjuku-ku, Tokyo 162-8601, Japan\\
$^3$Department of Physics, Waseda University, Shinjuku, 
Tokyo 169-8555, Japan}

\begin{abstract}

We study a coupled dark energy scenario in which a massive 
vector field $A_{\mu}$ with broken $U(1)$ gauge symmetry interacts 
with the four-velocity $u_c^{\mu}$ of cold dark matter (CDM) 
through the scalar product $Z=-u_c^{\mu} A_{\mu}$.
This new coupling corresponds to the momentum transfer, so 
that the background vector and CDM continuity equations 
do not have explicit interacting terms analogous to 
the energy exchange. Hence the observational preference 
of uncoupled generalized Proca theories over the $\Lambda$CDM model 
can be still maintained at the background level.
Meanwhile, the same coupling strongly affects the evolution 
of cosmological perturbations. 
While the effective sound speed of CDM vanishes, the propagation 
speed and no-ghost condition of a longitudinal scalar of $A_{\mu}$ 
and the CDM no-ghost condition are subject to nontrivial modifications 
by the $Z$ dependence in the Lagrangian. 
We propose a concrete dark energy model and show that the 
gravitational interaction on scales relevant to the linear 
growth of large-scale structures can be smaller than the Newton 
constant at low redshifts.
This leads to the suppression of growth rates of both CDM and total matter 
density perturbations, so our model allows an interesting possibility 
for reducing the tension of matter density contrast $\sigma_8$ 
between high- and low-redshift measurements.
 
\end{abstract}

\date{\today}

\pacs{04.50.Kd, 95.36.+x, 98.80.-k}

\maketitle

\section{Introduction}
\label{introsec}

The energy density of today's Universe is dominated by 
dark energy and dark matter, besides a small amount 
of baryons ($\sim$ 5\,\%). The standard paradigm of this
dark sector is known as the $\Lambda$CDM 
model \cite{Peebles1,Peebles2}, 
in which the origins of two dark components are a cosmological 
constant ($\Lambda$) and the cold dark matter (CDM). 
The cosmological constant is the simplest 
possibility for realizing late-time cosmic acceleration, 
but there has been a growing tension regarding today's Hubble expansion rate
$H_0$ between Cosmic Microwave 
Background (CMB) temperature anisotropies and low-redshift 
measurements \cite{Riess:2016jrr,Aghanim:2018eyx,Verde:2019ivm,Riess:2019cxk,Freedman:2019jwv,Wong:2019kwg,Reid:2019tiq}. 
Moreover, the observational data associated with galaxy clusterings and 
weak lensing typically favor the 
amplitude of matter density contrast $\sigma_8$ smaller than 
that constrained by CMB \cite{Macaulay:2013swa,Nesseris:2017vor,
Hildebrandt:2016iqg,Joudaki:2017zdt}.

The cosmological constant predicts a constant dark energy 
equation of state $w_{\rm DE}=-1$, but dynamical models of 
late-time cosmic acceleration generally lead to the time 
variation of $w_{\rm DE}$ \cite{CSW}. 
For example, a canonical scalar 
field dubbed quintessence \cite{quin1,quin2,quin3,quin4,quin5} 
gives rise to the time-varying 
$w_{\rm DE}$ in the range $w_{\rm DE}>-1$. 
However, there has been no significant observational evidence 
that quintessence is favored over the $\Lambda$CDM 
model \cite{CDT,Tsujikawa:2013fta}.
Meanwhile, the phantom equation of state ($w_{\rm DE}<-1$) 
allows a possibility for exhibiting better compatibility 
with the data in comparison to the $\Lambda$CDM model. 
In the presence of scalar or vector fields with derivative 
self-interactions or nonminimal couplings to gravity, 
it is possible to realize $w_{\rm DE}<-1$ without the 
appearance of ghosts \cite{Hu:2007nk,Tsujikawa:2008uc,DT10,DeFelice:2016yws}.

The gravitational-wave (GW) event GW170817 \cite{GW170817}, together with its 
electromagnetic counterparts \cite{Goldstein}, showed that the speed of gravity 
$c_T$ is very close to that of light $c$ in the redshift range $z<0.009$. 
If we strictly demand that $c_T=c$ without any tunings
among functions, a large set of nonminimal couplings to gravity 
are forbidden in scalar-tensor and vector-tensor theories \cite{Lon15,GWcon1,GWcon2,GWcon3,GWcon4,GWcon5,GWcon6}. 
In generalized Proca (GP) theories, which correspond
to vector-tensor theories with second-order equations 
of motion \cite{Heisenberg,Tasinato1,Tasinato2,Fleury,Hull,Allys,Jimenez16,Allys2}, 
the resulting action should contain the minimally 
coupled Ricci scalar $R$ and the Galileon-like Lagrangians up to 
cubic order, besides intrinsic vector modes \cite{Kunz}. 
Dark energy models in GP theories predict $w_{\rm DE}$ less than 
$-1$ in the matter era, which is followed by a self-accelerating de Sitter 
attractor with $w_{\rm DE}=-1$ \cite{DeFelice:2016yws,DeFelice:2016uil,deFelice:2017paw}.
At the background level, such models can show better compatibility 
with the current observational data in comparison to the $\Lambda$CDM 
model by reducing the tension 
of $H_0$ \cite{deFelice:2017paw,Nakamura:2018oyy,DeFelice:2020sdq}.

As for the evolution of cosmological perturbations relevant to 
galaxy clusterings, the cubic-order GP theories 
predict the effective gravitational coupling  $G_{\rm eff}$ with matter 
larger than the Newton constant $G$ \cite{DeFelice:2016uil,Kunz,Nakamura:2018oyy}. 
In this case, the growth of matter perturbations is enhanced by the 
cubic derivative coupling, so the $\sigma_8$ tension present in the $\Lambda$CDM model 
tends to get worse in general.
This also limits the compatibility of GP theories against cross-correlation data 
between the integrated Sachs-Wolfe (ISW) signal and the galaxy distribution.
Indeed, the Markov-chain-Monte-Carlo analysis of 
Ref.~\cite{Nakamura:2018oyy} showed that inclusion of the data of 
ISW-galaxy cross-correlations and redshift-space distortions 
does not improve constraints derived from the background 
expansion history. This situation is even severer in cubic-order 
scalar-tensor (Horndeski) theories \cite{Kobayashi:2009wr,Kimura:2011td}, for which
the absence of vector degrees of freedom does not render $G_{\rm eff}$ close to $G$.

If the vector field $A_{\mu}$ is coupled to CDM, there may be a possibility 
that the gravitational coupling with CDM is smaller than $G$. 
In Ref.~\cite{Nakamura2019}, the coupled dark energy scenario with the 
interacting Lagrangian ${\cal L}_{\rm int}=Q f(X) \rho_c$ was proposed, 
where $Q$ is a coupling constant, $f$ is a function of $X=-A^{\mu}A_{\mu}/2$, and $\rho_c$ is the CDM 
density (see also Ref.~\cite{Gomez}). 
This is analogous to the Lagrangian ${\cal L}_{\rm int}=Q \dot{\phi} \rho_c$ \cite{Wette,Amendola:1999er}
studied in the context of scalar-tensor theories, where $\dot{\phi}$ 
is the time derivative of scalar field $\phi$. 
These interactions correspond to the energy transfer, which 
typically works to enhance the gravitational coupling with CDM. 
In coupled quintessence, for example, the gravitational coupling with CDM is given by $G_{\rm eff}=(1+2Q^2) G$ \cite{Amendola:2003wa}. 

There is yet other kind of interactions associated with the momentum 
transfer. In scalar-tensor theories, the field-derivative coupling with the 
CDM four-velocity $u_c^{\mu}$, which is quantified by the scalar combination 
$Z=u_c^{\mu} \partial_{\mu} \phi$ \cite{Pourtsidou:2013nha,Boehmer:2015sha,Skordis:2015yra,
Dutta:2017kch}, can give rise to the CDM gravitational coupling smaller than 
$G$ \cite{Koivisto:2015qua,Pourtsidou:2016ico,Kase:2019veo,Kase:2019mox,Chamings:2019kcl,Amendola:2020ldb} on scales relevant to the linear growth of large-scale structures.
In GP theories, the interaction analogous to the momentum transfer 
in scalar-tensor theories is quantified by the scalar combination 
$Z=-u_c^{\mu}A_{\mu}$. 
The existence of intrinsic vector modes in GP theories generally affects the gravitational 
coupling with CDM \cite{DeFelice:2016uil,Nakamura:2018oyy}, and it has not been clarified yet
whether the weak cosmic growth can be realized in coupled GP theories with the momentum transfer.

To shed some light on this issue, in this paper, we study the cosmology of cubic-order GP theories with the 
interacting Lagrangian of the form $f(X,Z)$, where $f$ is a function of 
$X$ and $Z$. We consider the case in which the vector field is 
only coupled to CDM, but uncoupled to baryons or radiation. 
Then, there are no conflicts with local gravity experiments \cite{DeFelice:2016cri}.
The CDM, baryons, and radiation are assumed to be perfect fluids, 
which are described by a Schutz-Sorkin action \cite{Sorkin,Brown,SorkinADF}.
At the background level, the interacting terms do not explicitly appear 
on the right-hand-sides of vector-field and CDM continuity equations, 
so it is possible to maintain the good cosmological background known for 
uncoupled GP theories \cite{DeFelice:2016yws,deFelice:2017paw,Nakamura:2018oyy,DeFelice:2020sdq}. 
We also derive the general expression of 
effective gravitational couplings for CDM and baryon perturbations 
on scales deep inside the sound horizon. 
Finally, we propose a concrete coupled dark energy model with the 
explicit $Z$ dependence in the Lagrangian and show that the weak 
cosmic growth of both CDM and total matter density perturbations 
can be realized by the momentum exchange between the vector field and CDM.

Throughout the paper, we adopt the units for which the speed of light $c$,
the reduced Planck constant $\hbar$, and the Boltzmann 
constant $k_B$ are set to unity.
The reduced Planck mass $M_{\rm pl}$ is related to 
the Newton gravitational constant $G$, as 
$M_{\rm pl}^2=1/(8\pi G)$.
The Greek and Latin indices represent components in four-dimensional 
space-time and in a three-dimensional space, respectively.

\section{Coupled generalized Proca theories with momentum transfer}
\label{eomsec}

We consider cubic-order GP theories with a vector field $A_{\mu}$. 
The vector field breaks a $U(1)$ gauge symmetry due to the
existence of Lagrangians $G_2(X)$ and $G_3(X) \nabla_{\mu}A^{\mu}$, 
where $G_2$ and $G_3$ are functions of $X=-A^{\mu}A_{\mu}/2$ and 
$\nabla_{\mu}$ is the covariant derivative operator. 
In this case, the vector field can play a role of dark energy with late-time 
cosmic acceleration \cite{DeFelice:2016yws,DeFelice:2016uil,deFelice:2017paw}.
We assume that CDM is described by a perfect fluid 
with the four-velocity $u_{c}^{\mu}$. 
Given the unknown properties of dark sectors, 
we would like to consider possible interactions between them 
which are present at the level of Lagrangian. 
In coupled GP theories, there exists a simple interaction
quantified by a scalar combination, 
\be
Z=-u_{c}^{\mu}A_{\mu}\,.
\ee
As we will explicitly show in this paper, 
this new coupling allows a possibility for realizing the weak cosmic 
growth. Whether or not this type of coupling can arise from some 
fundamental particle theories is an open question, 
which deserves for a future study.

The action of our coupled GP theories is given by 
\be
{\cal S}= \int {\rm d}^{4}x \sqrt{-g} 
\left[ \frac{M_{\rm pl}^2}{2} R-\frac{1}{4} 
F_{\mu \nu} F^{\mu \nu}
+f \left( X,Z \right)
+G_3(X) \nabla_{\mu}A^{\mu}\right]
+{\cal S}_M\,,
\label{action}
\ee
where $g$ is the determinant of metric tensor 
$g_{\mu\nu}$, $R$ is the Ricci scalar, 
and $F_{\mu \nu}=\nabla_{\mu} A_{\nu}-\nabla_{\nu} A_{\mu}$. 
The function $f$, which is the generalization of $G_2(X)$,  
depends on both $X$ and $Z$.
For the matter action ${\cal S}_M$, we consider 
the perfect fluids of CDM, baryons, and radiation, which are 
labelled by $I=c,b,r$, respectively.
The perfect fluids can be described by 
the Schutz-Sorkin action\footnote{An equivalent action with respect to a four 
vector instead of the vector density $J_I^\mu$ has been introduced in Ref.~\cite{SorkinADF}.} \cite{Sorkin,Brown},
\be
{\cal S}_{M} = - \sum_{I=c,b,r}\int {\rm d}^{4}x 
\left[ \sqrt{-g}\,\rho_I(n_I)
+ J_I^{\mu} \left( \partial_{\mu} \ell_I 
+ \mathcal{A}_{I1} \partial_{\mu} \mathcal{B}_{I1} 
+ \mathcal{A}_{I2} \partial_{\mu} \mathcal{B}_{I2} 
\right) \right]\,,
\label{Schutz}
\ee
where the operator $\partial_{\mu}$ represents 
the partial derivative with respect to the coordinate $x^{\mu}$.
The fluid density $\rho_I$ depends on its number density $n_I$, 
which is related to the vector field $J_I^{\mu}$, as
\be
n_I=\sqrt{\frac{J_I^{\mu} J_I^{\nu}
g_{\mu \nu}}{g}}\,.
\label{ndef}
\ee
The scalar quantity $\ell_I$ is a Lagrange multiplier, whose variation 
leads to a constraint of the particle number conservation.
The quantities $\mathcal{A}_{I1}$, $\mathcal{A}_{I2}$ and 
$\mathcal{B}_{I1}$, $\mathcal{B}_{I2}$ 
are the Lagrange multipliers and Lagrange coordinates of fluids, respectively, 
both of which can be regarded as the two components of 
spatial vector fields $\mathcal{A}_{Ij}$ and $\mathcal{B}_{Ij}$ 
($j=1,2,3$). Since these fields are associated with intrinsic 
vector modes, the divergence-free conditions 
give the two independent components $\mathcal{A}_{I1}$, $\mathcal{A}_{I2}$ and 
$\mathcal{B}_{I1}$, $\mathcal{B}_{I2}$ for each of them.
Since there exists a dynamical vector field in GP theories, 
we need to take the Lagrangian $-J_I^{\mu}(\mathcal{A}_{I1} \partial_{\mu} \mathcal{B}_{I1} 
+ \mathcal{A}_{I2} \partial_{\mu} \mathcal{B}_{I2})$ into account 
for the analysis of vector perturbations \cite{DeFelice:2016yws,DeFelice:2016uil}.
In Sec.~\ref{vectorsec}, we will study the dynamics of vector perturbations 
by varying the action (\ref{Schutz}) with respect to $\mathcal{A}_{I1}$, $\mathcal{A}_{I2}$, 
$\mathcal{B}_{I1}$, $\mathcal{B}_{I2}$.

The fluid four-velocity $u_{I{\mu}}$ is defined by 
\be
u_{I{\mu}}=\frac{J_{I{\mu}}}{n_I\sqrt{-g}}\,,
\label{udef}
\ee
which obeys $u_I^{\mu} u_{I{\mu}}=-1$ from Eq.~(\ref{ndef}). 
The scalar combination $Z$ is expressed as 
\be
Z=-\frac{g^{\mu \nu} J_{c\mu} A_{\nu}}{n_c\sqrt{-g}}\,.
\label{Z}
\ee
Neither radiation nor baryons are assumed to be 
coupled to the vector field.

\subsection{Covariant equations of motion}

We derive the covariant equations of motion by varying 
(\ref{action}) with respect to several variables in the action. 
Variation with respect to $\ell_I$ leads to  
\be
\partial_{\mu} J_I^{\mu}=0\,,
\label{Jmure}
\ee
which holds for each $I=c,b,r$.
On using the property $J_I^{\mu}=n_I \sqrt{-g}\,u_I^{\mu}$ 
and the relation $\partial_{\mu} (\sqrt{-g} u_I^{\mu})=
\sqrt{-g} \nabla_{\mu} u_I^{\mu}$, 
Eq.~(\ref{Jmure}) translates to 
\be
n_I \nabla_{\mu} u_I^{\mu}+u_I^{\mu} 
\partial_{\mu} n_I=0\,.
\label{nIre}
\ee
Since $\rho_I$ depends only on $n_I$, there is  
the relation,
\be
\left( \rho_I+P_I \right) \partial_{\mu} n_I=
n_I \partial_{\mu} \rho_I\,,
\label{rhoIn}
\ee
where $P_I$ is the fluid pressure defined by 
\be
P_I=n_I \rho_{I,n_I}-\rho_I\,,
\label{Pdef}
\ee
with the notation 
$\rho_{I,n_I} \equiv \partial \rho_I/\partial n_I$.
On using Eqs.~(\ref{nIre}) and (\ref{rhoIn}), we obtain 
\be
u_I^{\mu} \partial_{\mu} \rho_I+
\left( \rho_I+P_I \right) 
\nabla_{\mu} u_I^{\mu}=0\,.
\label{conser1}
\ee
We vary the action (\ref{action}) with respect to $J_c^{\mu}$
by keeping in mind that the scalar combination $Z$ of Eq.~(\ref{Z}) 
depends on $J_c^{\mu}$. On using the property 
$\partial n_I/\partial J_I^{\mu}=J_{I{\mu}}/(n_I g)$, 
it follows that 
\be
\partial_{\mu} \ell_c= 
u_{c {\mu}} \rho_{c,n_c}
-\frac{f_{,Z}}{n_c} \left( A_{\mu}-Z u_{c \mu} 
\right)
-\mathcal{A}_{c1} \partial_{\mu} \mathcal{B}_{c1} 
-\mathcal{A}_{c2} \partial_{\mu} \mathcal{B}_{c2}
\,.
\label{lc}
\ee
For baryons and radiation, there is no dependence of 
$J_b^{\mu}$ and $J_r^{\mu}$ in the function $f$, 
so that 
\be
\partial_{\mu} \ell_I= 
u_{I {\mu}} \rho_{I,n_I}
-\mathcal{A}_{I1} \partial_{\mu} \mathcal{B}_{I1} 
-\mathcal{A}_{I2} \partial_{\mu} \mathcal{B}_{I2}\,,
\label{lbr}
\ee
where $I=b,r$.

The covariant Einstein equations of motion follow by varying the action 
(\ref{action}) with respect to $g^{\mu \nu}$. 
In doing so, we use the following properties,
\ba
\delta n_I &=& \frac{n_I}{2} \left( g_{\mu \nu} 
-u_{I\mu} u_{I\nu} \right) \delta g^{\mu \nu}\,,\\
\delta X &=&-\frac{1}{2} A_{\mu} A_{\nu}
\delta g^{\mu \nu}\,,\\
\delta Z &=& \left( \frac{1}{2} Z u_{c\mu} 
u_{c\nu}-u_{c\mu} A_{\nu} \right) 
\delta g^{\mu \nu}\,,
\ea
together with 
$\delta \sqrt{-g}=-(1/2)\sqrt{-g} g_{\mu \nu} \delta g^{\mu \nu}$. 
Then, the resulting covariant equations are given by 
\be
M_{\rm pl}^2 G_{\mu\nu}=\sum_{I=c,b,r} T^{(I)}_{\mu\nu}
+T_{\mu\nu}^{(A)}\,,
\label{Ein}
\ee
where $G_{\mu \nu}$ is the Einstein tensor, and
\ba
T^{(I)}_{\mu\nu} 
&=& 
\left( \rho_I+P_I \right) 
u_{I \mu} u_{I \nu}+P_I g_{\mu \nu} \,,
\label{Tmmun}\\
T^{(A)}_{\mu \nu} 
&=& F_{\mu \rho} {F_{\nu}}^{\rho} -\frac{1}{4} 
g_{\mu \nu} F_{\rho \sigma} F^{\rho \sigma}
+f g_{\mu \nu}
+f_{,X} A_{\mu} A_{\nu}+f_{,Z} Z u_{c\mu} u_{c\nu} 
\nonumber \\
& &
+G_{3,X} \left( A_{\mu} A_{\nu} \nabla_{\rho}A^{\rho}
+g_{\mu \nu} A^{\lambda} A_{\rho} \nabla_{\lambda} A^{\rho}
-A_{\rho}  A_{\mu} \nabla_{\nu} A^{\rho}
-A_{\rho}  A_{\nu} \nabla_{\mu} A^{\rho} \right)\,.
\label{Tmmund}
\ea
Varying the action (\ref{action}) with respect to $A_{\nu}$, 
the equation for the vector field yields
\be
\nabla_{\mu} F^{\mu \nu}-f_{,X}A^{\nu}-f_{,Z} u_{c}^{\nu} 
+G_{3,X} \left( A^{\mu} \nabla^{\nu} A_{\mu}
-A^{\nu} \nabla^{\mu} A_{\mu} \right)=0\,.
\label{Amueq}
\ee

Taking the covariant derivative of Eq.~(\ref{Ein}) 
leads to
\be
\sum_{I=c,b,r} \nabla^{\mu} T^{(I)}_{\mu\nu}
+\nabla^{\mu} T_{\mu \nu}^{(A)}=0\,.
\label{Tcon}
\ee
On using the property (\ref{conser1}), it follows that 
\be
u_I^{\nu} \nabla^{\mu} T^{(I)}_{\mu\nu} =0\,,
\label{Tcon1}
\ee
which holds for $I=c,b,r$.
This corresponds to the continuity equation for each 
perfect fluid.
If CDM is the only fluid component, we have
$u_c^{\nu} \nabla^{\mu} T^{(A)}_{\mu\nu}
=-u_c^{\nu} \nabla^{\mu} T^{(c)}_{\mu\nu}=0$ 
from Eqs.~(\ref{Tcon}) and (\ref{Tcon1}).
Since we are considering coupled GP theories 
with the momentum transfer alone, there are no explicit 
interacting terms associated with the energy exchange.
This property is different from interacting GP theories 
with the energy transfer studied in Ref.~\cite{Nakamura2019}.
We note that the momentum exchange between the vector 
field and CDM occurs through Eq.~(\ref{Tcon}).

\subsection{Background equations of motion}

We derive the background equations on the flat 
Friedmann-Lema\^itre-Robertson-Walker (FLRW) 
spacetime given by the line element, 
\be
{\rm d}s^2=-{\rm d}t^2+a^2(t) \delta_{ij} {\rm d}x^i {\rm d} x^j\,,
\label{metric}
\ee
where $a$ is the scale factor that depends on the cosmic 
time $t$. The vector-field profile and the fluid four-velocities 
consistent with this background are given, respectively, by 
\be
A^{\mu}=\left( \phi(t), 0, 0, 0 \right)\,,\qquad 
u_I^{\mu}=\left( 1, 0, 0, 0 \right)\,,
\ee
where $\phi$ is a function of $t$. 
We introduce the Hubble-Lema\^itre expansion rate 
$H=\dot{a}/a$, where a dot denotes
a derivative with respect to $t$. 
Since $\nabla_{\mu} u_I^{\mu}=3H$, the fluid continuity 
Eq.~(\ref{Tcon1}), which is equivalent to Eq.~(\ref{conser1}), 
reduces to 
\be
\dot{\rho}_I+3H \left( \rho_I+P_I \right)=0\,,
\label{conFLRW}
\ee
with $I=c, b, r$.

{}From the (00) and $(ii)$ components of Einstein equations 
(\ref{Ein}), we obtain  
\ba
& & 
3M_{\rm pl}^2 H^2=\sum_{I=c,b,r} \rho_I
-f+\left( f_{,X}\phi+f_{,Z}+3G_{3,X}H \phi^2 
\right) \phi\,,
\label{Eq00}\\
& & 
M_{\rm pl}^2 \left( 2\dot{H}+3H^2 
\right)=-\sum_{I=c,b,r}P_I-f+G_{3,X}\phi^2 \dot{\phi}\,.
\label{Eq11}
\ea
The $\nu=0$ component of Eq.~(\ref{Amueq}) 
translates to
\be
f_{,X}\phi+f_{,Z}+3G_{3,X}H \phi^2=0\,.
\label{Aeq}
\ee
We define the dark energy density $\rho_{\rm DE}$ and 
pressure $P_{\rm DE}$, as 
\ba
\rho_{\rm DE} 
&=& -f+\left( f_{,X}\phi+f_{,Z}+3G_{3,X}H \phi^2 
\right) \phi=-f\,,\label{rhode}\\
P_{\rm DE}
&=& f-G_{3,X} \phi^2 \dot{\phi}\,,
\ea
where we used Eq.~(\ref{Aeq}) in the second equality of 
Eq.~(\ref{rhode}). 
Taking the time derivative of Eq.~(\ref{rhode}) and 
exploiting Eq.~(\ref{Aeq}), we obtain 
\be
\dot{\rho}_{\rm DE}+3H \left( \rho_{\rm DE}
+P_{\rm DE} \right)=0\,,
\label{rhoDEeq}
\ee
which corresponds to the continuity equation
in the dark energy sector.

Taking the time derivative of Eq.~(\ref{Aeq}) and combining it with 
Eq.~(\ref{Eq11}), it follows that 
\ba
\dot{\phi} &=& \frac{\phi^4 G_{3,X}}{q_S} 
\left( 3\rho_c+3\rho_b+4 \rho_r \right)\,,\label{dotphi}\\
\dot{H} &=& -\frac{q_S-3\phi^6 G_{3,X}^2}
{6M_{\rm pl}^2 q_S}
\left( 3\rho_c+3\rho_b+4 \rho_r \right)\,,
\label{dotH}
\ea
where 
\be
q_S=3\phi^3 \left( 2H \phi^2 M_{\rm pl}^2 G_{3,XX}
+\phi^3 G_{3,X}^2+4HM_{\rm pl}^2 G_{3,X} \right)
+2\phi^2 M_{\rm pl}^2 \left( \phi^2 f_{,XX}
+2\phi f_{,XZ}+f_{,ZZ}+f_{,X} \right)\,.
\label{qS0}
\ee
As we will show later in Sec.~\ref{sec3}, the quantity $q_S$ 
must be positive to avoid the ghost in the scalar sector. 
In this case, the right hand sides of Eqs.~(\ref{dotphi}) and (\ref{dotH}) 
do not cross the singular point $q_S=0$.
 
We also introduce the density parameters,
\be
\Omega_I=\frac{\rho_{I}}{3M_{\rm pl}^2 H^2}\,,
\qquad 
\Omega_{\rm DE} = \frac{\rho_{\rm DE}}{3M_{\rm pl}^2 H^2}\,.
\label{Omega}
\ee
as well as the equations of state 
\be
w_I=\frac{P_I}{\rho_I}\,,\qquad 
w_{\rm DE}=\frac{P_{\rm DE}}{\rho_{\rm DE}}
=-1+\frac{G_{3,X} \phi^2 \dot{\phi}}{f}\,. 
\label{wde}
\ee
Then, Eq.~(\ref{Eq00}) is expressed as
\be
\sum_{I=c,b,r} \Omega_I+\Omega_{\rm DE}=1\,.
\label{tOmega}
\ee
The effective equation of state is given by 
\be
w_{\rm eff}=\sum_{I=c,b,r} w_I \Omega_I+
w_{{\rm DE}} \Omega_{{\rm DE}}= -1-\frac{2\dot{H}}{3H^2}\,,
\label{weff}
\ee
where we used Eq.~(\ref{Eq11}) in the second equality. 
The $Z$ dependence in $f$ affects the evolution of $\phi$ 
through the term $f_{,Z}$ in Eq.~(\ref{Aeq}).
The dark energy equation of state $w_{\rm DE}$ is 
also modified by the vector-CDM interaction.

\section{Cosmological perturbations and theoretically 
consistent conditions}
\label{sec3}

We proceed to the study of cosmological perturbations on the flat FLRW background 
(\ref{metric}). The linear perturbations can be decomposed into tensor, vector, and 
scalar modes, which evolve independently from each other. 
The perturbed line element in the flat gauge is given by 
\be
{\rm d}s^{2} = - \left( 1 + 2 \alpha \right) {\rm d}t^{2}
+ 2 \left( \partial_{i}\chi + V_{i} \right) {\rm d}t\,{\rm d}x^{i}
+ a^{2}(t) \left( \delta_{ij} + h_{ij} \right) {\rm d}x^{i} {\rm d}x^{j}\,,
\label{permet}
\ee
where $\alpha$ and $\chi$ are scalar perturbations with the notation 
$ \partial_{i}\chi =\partial \chi/\partial x^i$, $V_i$ is the vector perturbation 
obeying the transverse condition $\partial^i V_i=0$, and $h_{ij}$ is the tensor 
perturbation satisfying the transverse and traceless conditions $\partial^i h_{ij}=0$ and $h^i_i=0$. 

The vector field $J_I^{\mu}$ in the Schutz-Sorkin action (\ref{Schutz}) 
contains both scalar and vector modes, such that 
\be
J_I^{0}={\cal N}_I+\delta J_I\,,\qquad
J_I^{i}= \frac{1}{a^2(t)} \delta^{ij} 
\left( \partial_j \delta j_I+W_{Ij} \right)\,,
\ee
where $\delta J_I$ and $\delta j_I$ are scalar perturbations, and $W_{Ij}$
is the vector perturbation satisfying $\partial^j W_{Ij}=0$. 
Here, ${\cal N}_I$ is the background particle number of each matter species, which 
is constant from Eq.~(\ref{Jmure}).
We also decompose the vector field $A^{\mu}$, as 
\be
A^0=\phi(t)+\delta \phi\,,\qquad 
A^i=\frac{1}{a^{2}(t)} \delta^{ij} 
\left( \partial_{j} \chi_{V} + E_{j}\right)\,,
\label{A0i}
\ee
where $\delta \phi$ and $\chi_{V}$ are scalar perturbations, and $E_{j}$
is the vector perturbation satisfying $\partial^j E_{j}=0$. 
Substituting $g_{0i}=\partial_{i}\chi + V_{i}$, $g_{ij}=a^2(t)\delta_{ij}$, 
and Eq.~(\ref{A0i}) into $A_i=g_{0i} A^0+g_{ij}A^j$, 
the spatial component of $A_{\mu}$ yields
\be
A_i=\partial_i \psi+Y_i\,,
\label{Aid}
\ee
where 
\ba
& &
\psi \equiv \chi_V+\phi(t) \chi\,,\\
& &
Y_{i} \equiv E_{i} + \phi(t) V_{i}\,.
\ea
The perturbations $\psi$ and $Y_i$ correspond to the dynamical 
scalar and vector degrees of freedom, respectively.

The spatial component of $u_{I \mu}$  
can be expressed in the form 
\be
u_{Ii} = -\partial_{i}v_{I} + v_{Ii}\,,
\label{u_i}
\ee
where $v_I$ is the scalar velocity potential, and $v_{Ii}$ is 
the intrinsic vector mode satisfying $\partial^i v_{Ii}=0$.

Substituting Eqs.~(\ref{Aid}) and (\ref{u_i}) into the spatial component of 
Eq.~(\ref{lc}), it follows that 
\be
\partial_i \ell_c 
+\mathcal{A}_{c1} \partial_{i} \mathcal{B}_{c1} 
+\mathcal{A}_{c2} \partial_{i} \mathcal{B}_{c2} 
= -\rho_{c,n_c} \partial_i v_c
-\frac{f_{,Z}}{n_c} \left( \partial_i \psi+\phi\,
\partial_i v_c \right) +\rho_{c,n_c} v_{ci}-\frac{f_{,Z}}{n_c} 
\left( Y_i -\phi\,v_{ci} \right)\,,
\label{lcex}
\ee
up to linear order in perturbations. The coefficients in front of 
the perturbed quantities in Eq.~(\ref{lcex}) (e.g., $\rho_{c,n_c}$)  are 
time-dependent background quantities. 
The rotational-free scalar part $\partial_i \ell_c$ 
needs to be identical to the spatial derivative of scalar perturbations 
on the right-hand-side of Eq.~(\ref{lcex}), while the divergence-free
vector part $\mathcal{A}_{c1} \partial_{i} \mathcal{B}_{c1} 
+\mathcal{A}_{c2} \partial_{i} \mathcal{B}_{c2}$ is equivalent to 
the corresponding intrinsic vector perturbations on the same 
right-hand-side.
This gives the following relations, 
\ba
& &
\partial_i \ell_c 
= -\rho_{c,n_c} \partial_i v_c
-\frac{f_{,Z}}{n_c} \left( \partial_i \psi+\phi\,\partial_i v_c 
\right)\,,\label{la1}\\
& & \mathcal{A}_{c1} \partial_i \mathcal{B}_{c1}
+\mathcal{A}_{c2} \partial_i \mathcal{B}_{c2}
=\rho_{c,n_c} v_{ci}-\frac{f_{,Z}}{n_c} 
\left( Y_i -\phi\,v_{ci} \right)\,.
\label{la2}
\ea
The integrated solution to Eq.~(\ref{la1}) is
$\ell_c=c(t)-\rho_{c,n_c} v_c
-(f_{,Z}/n_c) \left( \psi+\phi\,v_c \right)$. 
The time-dependent function $c(t)$ is determined by the 
$\mu=0$ component of Eq.~(\ref{lc}), as 
$c(t)=-\int^t \rho_{c,n_c}(\tilde{t}){\rm d} \tilde{t}$. 
Then, the scalar quantity $\ell_c$ is given by 
\be
\ell_c=-\int^t \rho_{c,n_c}(\tilde{t}){\rm d} \tilde{t}
-\rho_{c,n_c} v_c
-\frac{f_{,Z}}{n_c} \left( \psi+\phi\, v_c 
\right)\,,
\label{lcf}
\ee
which contains the velocity potential $v_c$ and the 
dynamical perturbation $\psi$.
We recall that the energy-momentum tensors (\ref{Tmmun}) and 
(\ref{Tmmund}) were obtained after eliminating $\ell_c$ on 
account of Eq.~(\ref{lc}). 
The terms $-\rho_{c,n_c} v_c$ and 
$-(f_{,Z}/n_c) \left( \psi+\phi\, v_c \right)$ in Eq.~(\ref{lcf}) 
contribute to Eqs.~(\ref{Tmmun}) and (\ref{Tmmund}), respectively, as 
the perturbed energy-momentum tensors.

Since the linear perturbations with different 
wave numbers do not mix on the FLRW background,  
we can consider a configuration with which all the perturbations 
propagate in one direction, $x_3$. 
Then, the vector perturbations $X_i=V_i, W_{Ii}, E_i, v_{ci}$ depend on 
$t$ and $x_3$. The components of $X_i$ consistent with 
the divergence-free conditions $\partial^i X_i=0$ are chosen to be
\be
X_{i} = \left(X_{1}(t,x_3),\, X_{2}(t,x_3),\,0\right)\,.
\ee
For the Lagrange multiplers ${\cal A}_{I1}$, ${\cal A}_{I2}$, ${\cal B}_{I1}$,  
${\cal B}_{I2}$, we can choose them in the following forms \cite{DeFelice:2009bx}
\ba
& &
{\cal A}_{I1} = \delta{\cal A}_{I1}(t,x_3)\,,\qquad 
{\cal A}_{I2} = \delta{\cal A}_{I2}(t,x_3)\,, \\
& &
{\cal B}_{I1} = x_1 + \delta{\cal B}_{I1}(t,x_3)\,, \qquad
{\cal B}_{I2} = x_2 + \delta{\cal B}_{I2}(t,x_3)\,,
\label{dBdef}
\ea
where $\delta{\cal A}_{I1}$, $\delta{\cal A}_{I2}$, 
$\delta{\cal B}_{I1}$, $\delta{\cal B}_{I2}$ are perturbed 
quantities. The vector perturbations 
$\delta{\cal A}_{Ii}=(\delta{\cal A}_{I1}(t,x_3), \delta{\cal A}_{I2}(t,x_3), 0)$ 
and 
$\delta{\cal B}_{Ii}=(\delta{\cal B}_{I1}(t,x_3), \delta{\cal B}_{I2}(t,x_3), 0)$ 
satisfy the transverse conditions 
$\partial^i \delta{\cal A}_{Ii}=0$ and $\partial^i \delta{\cal B}_{Ii}=0$. 
The vector field ${\cal B}_{Ii}$, which is orthogonal to the $x_3$ direction, 
can be chosen to have the background components 
${\bar {\cal B}}_{Ii}=(b_1 x_1, b_2 x_2, 0)$ with arbitrary constants 
$b_1$ and $b_2$. 
In Eq.~(\ref{dBdef}) both $b_1$ and $b_2$ are normalized to be 1, 
in which case the left-hand side of Eq.~(\ref{la2}) reduces to 
the linear perturbation $\delta {\cal A}_{ci}$ (with $i=1,2$). 
This is consistent with the fact that the right-hand-side of 
Eq.~(\ref{la2})  consists of the perturbations at linear order.
Then, it follows that 
\be
\delta {\cal A}_{ci}=\rho_{c,n_c} v_{ci}-\frac{f_{,Z}}{n_c} 
\left( Y_i -\phi\,v_{ci} \right)\,.
\label{dlAc}
\ee
On using Eq.~(\ref{lbr}), the relations for baryons and radiation 
analogous to Eqs.~(\ref{lcf}) and (\ref{dlAc}) are 
given, respectively, by
\ba
\ell_I &=& -\int^t \rho_{I,n_I}(\tilde{t}){\rm d} \tilde{t}
-\rho_{I,n_I} v_I\,,\label{lIf}\\
\delta {\cal A}_{Ii} &=&
\rho_{I,n_I} v_{Ii}\,,
\label{dlAI}
\ea
where $I=b,r$.

\subsection{Tensor perturbations}

The tensor perturbations $h_{ij}$, which are transverse and traceless,  
can be expressed in terms of the sum of two polarization modes, as 
$h_{ij} = h_{+}e_{ij}^{+} + h_{\times} e_{ij}^{\times}$. 
The unit vectors $e_{ij}^{+}$ and $e_{ij}^{\times}$ 
satisfy the normalizations 
$e_{ij}^{+}(\bm{k})e_{ij}^{+}(-\bm{k})^{*}=1$, 
$e_{ij}^{\times}(\bm{k})e_{ij}^{\times}(-\bm{k})^{*}=1$, and 
$e_{ij}^{+}({\bm k}) e_{ij}^{\times}(-{\bm k})^{*} = 0$ 
in Fourier space with the comoving wavenumber $\bm{k}$.
Expanding (\ref{action}) up to quadratic order in $h_{\lambda}$ 
(where $\lambda = +,\times$), integrating the action 
by parts, and using the background Eq.~(\ref{Eq11}), 
we end up with the second-order action 
of tensor perturbations,
\be
{\cal S}_{T}^{(2)} = \sum_{\lambda = +,\times}
\int {\rm d}t {\rm d}^{3}x\, \frac{M_{\rm pl}^2}{8} a^{3} \left[
\dot{h}_{\lambda}^{2} - \frac{1}{a^{2}} (\partial h_{\lambda})^{2}
\right] \,.
\ee
This is equivalent to the corresponding action of tensor perturbations 
in standard general relativity, 
so the speed of gravitational waves $c_T$ is equivalent to that of light.
Hence our coupled GP theories are consistent with 
the bound of $c_T$ constrained by the GW170817 event \cite{GW170817}.

\subsection{Vector perturbations}
\label{vectorsec}

The intrinsic vector modes appear in each term of (\ref{action}), 
so we sum up all those contributions to the action.
For this purpose, we use the fact that $\ell_I$ ($I=c,b,r$) are 
scalar quantities satisfying Eqs.~(\ref{lcf}) and (\ref{lIf}), so 
the term $J_I^{\mu} \partial_{\mu} \ell_I$ in the matter action 
(\ref{Schutz}) does not contribute to the quadratic-order 
action of vector perturbations.
Vary the resulting second-order action with respect to 
$W_{Ii}$ and $\delta {\cal A}_{Ii}$, it follows that 
\ba
W_{Ii} &=&\left( \frac{\delta {\cal A}_{Ii}}{\rho_{I,n_I}}-V_i \right){\cal N}_i \,,
\label{Wi} \\
\delta {\cal A}_{Ii} &=& \rho_{I,n_I}
\left( V_i-a^2 \dot{\delta {\cal B}}_{Ii} \right)\,.
\label{calAi}
\ea
The perturbations $\delta {\cal A}_{ci}$ and $\delta {\cal A}_{Ii}$ ($I=b,r$) 
are related to the spatial components of four-velocities 
according to Eqs.~(\ref{dlAc}) and (\ref{dlAI}), respectively.
Then, we have 
\ba
& &
V_i-a^2 \dot{\delta {\cal B}}_{ci}=v_{ci}-\frac{f_{,Z}}
{\rho_c+P_c} \left( Y_i -\phi\,v_{ci} \right)\,,
\label{velo1}\\
&& 
V_i-a^2 \dot{\delta {\cal B}}_{Ii}=v_{Ii}\,,\qquad 
({\rm for}~I=b,r)\,,
\label{velo2}
\ea
where we used Eq.~(\ref{Pdef}). 
In the following, we exploit Eqs.~(\ref{Wi}) and (\ref{calAi}) to eliminate the 
variables $W_{Ii}$ and $\delta {\cal A}_{Ii}$ from the second-order action.  
On using the background Eqs.~(\ref{Eq00}) and (\ref{Aeq}), 
the second-order action of vector perturbations yields
\ba
{\cal S}_{V}^{(2)} &=&
\int {\rm d}t {\rm d}^{3}x\, \sum_{i=1}^{2} \frac{a}{2} 
\biggl[
\dot{Y}_{i}^{2} - \frac{1}{a^{2}} (\partial Y_{i})^{2}
-\frac{1}{\phi} \left( G_{3,X} \phi \dot{\phi}-f_{,Z} \right) Y_{i}^{2}
-2f_{,Z} V_i Y_i
+ \frac{M_{\rm pl}^2}{2 a^{2}} (\partial V_{i})^{2}
\nonumber \\
& &
+ (V_i-a^2 \dot{\delta {\cal B}}_{ci})^2 (\rho_{c} + P_{c}+\phi f_{,Z}) 
+2a^2 f_{,Z} Y_i \dot{\delta {\cal B}}_{ci}
+\sum_{I=b,r}  (V_i-a^2 \dot{\delta {\cal B}}_{Ii})^2 (\rho_{I} + P_{I}) 
\biggr]\,.
\label{SV2}
\ea

In Fourier space with the comoving wavenumber $k=|{\bm k}|$, 
we vary the action (\ref{SV2}) with respect to $V_i$, $\delta {\cal B}_{ci}$, 
and $\delta {\cal B}_{Ii}$ ($I=b,r$). 
This leads to 
\ba
& &
\frac{M_{\rm pl}^2 k^2}{2a^2} V_i+\left( \rho_c+P_c+\phi f_{,Z} \right) 
\left( V_i-a^2  \dot{\delta {\cal B}}_{ci} \right)-f_{,Z}Y_i
+\sum_{I=b,r} \left( \rho_I+P_I \right) 
\left( V_i-a^2  \dot{\delta {\cal B}}_{Ii} \right)=0\,,
\label{vec1}\\
& &
\left[ \left( \rho_c+P_c+\phi f_{,Z} \right) 
\left( V_i-a^2  \dot{\delta {\cal B}}_{ci} \right)-f_{,Z}Y_i \right]a^3={\cal C}_{ci}\,,
\label{vec2}\\
& &
\left( \rho_I+P_I\right) 
\left( V_i-a^2  \dot{\delta {\cal B}}_{Ii} \right) a^3={\cal C}_{Ii}\,,\qquad
({\rm for}~I=b,r),
\label{vec3}
\ea
where ${\cal C}_{Ii}$ (with $I=c,b,r$) are constants in time. 
Notice that all the combinations in the form 
$V_i-a^2  \dot{\delta {\cal B}}_{Ii}$ (with $I=c,b,r$) can be rewritten in terms of the perfect fluid and Proca physical quantities by means of Eqs.\ (\ref{velo1}) and~(\ref{velo2}).
Substituting Eqs.~(\ref{vec2}) and (\ref{vec3}) into Eq.~(\ref{vec1}), we obtain
\be
V_i=-\frac{2}{M_{\rm pl}^2 k^2 a} \sum_{I=c,b,r} {\cal C}_{Ii}\,,
\label{vec4}
\ee
which decays as $|V_i| \propto a^{-1}$.
Plugging Eqs.~(\ref{velo1}) and (\ref{velo2}) into 
Eqs.~(\ref{vec2}) and (\ref{vec3}), it follows that 
\ba
v_{ci} &=&\frac{(\rho_c+P_c){\cal C}_{ci}+[2(\rho_c+P_c)
+\phi f_{,Z}]f_{,Z}
a^3 Y_i}{(\rho_c+P_c+\phi f_{,Z})^2 a^3}\,,
\label{vcir}\\
v_{Ii} &=& \frac{{\cal C}_{Ii}}{(\rho_I+P_I)a^3}\,,\qquad 
({\rm for}~I=b,r).
\ea
While $v_{bi}$ stays constant, the CDM velocity $v_{ci}$ 
is instead affected by the dynamical field $Y_i$.

Integrating out the Lagrange multiplier $V_i$ by means of Eq.~(\ref{vec1}), 
the action gets its reduced form, with the field $Y_i$ and 
the contributions from $\dot{\delta {\cal B}}_{ci}$, 
and $\dot{\delta {\cal B}}_{Ii}$ ($I=b,r$).
On taking the small-scale limit $k \to \infty$, 
the dominant contributions to the second-order 
action of vector perturbations are given by 
\be
{\cal S}_{V}^{(2)} \simeq \sum_{i=1}^{2} 
\int {\rm d}t{\rm d}^{3}x\,\frac{a}2
\left\{
 q_{V} \left[
\dot{Y}_{i}^{2} - c_{V}^{2} \frac{k^{2}}{a^{2}} Y_{i}^{2} 
\right]
+(\rho_c+P_c+\phi f_{,Z})\,a^4\,\dot{\delta {\cal B}}_{ci}^2
+\sum_{I=b,r}(\rho_I+P_I)\,a^4\,\dot{\delta {\cal B}}_{Ii}^2
\right\},
\label{SVf}
\ee
where
\be
q_{V} = 1 \,, \qquad 
c_{V}^{2} = 1\,.
\ee
Hence there are neither ghosts nor Laplacian instabilities for the dynamical perturbations 
$Y_i$, with the propagating speed equivalent to that of light. 
As we are going to see in Sec.~\ref{ssec}, the same no-ghost condition for the field 
$\delta {\cal B}_{ci}$, will reappear in the scalar perturbation sector, 
so that we will postpone its study for later.
Since the instability of $Y_i$ is absent, the violent growth of $v_{ci}$ 
does not occur through Eq.~(\ref{vcir}). 
This is the same conclusion as that found for uncoupled 
GP theories \cite{DeFelice:2016uil}. 
Hence the existence of dynamical vector perturbations does not affect 
the anisotropy in structure formation.
The constant $q_V$ different from 1 arises for
more general Lagrangians containing intrinsic vector modes, 
say, ${\cal L}_F=-q_V F_{\mu \nu}F^{\mu \nu}/4$.

The above discussion shows that the new interaction associated with the momentum transfer 
affects the small-scale stability conditions of neither tensor nor 
for the Proca vector perturbations.

\subsection{Scalar perturbations}
\label{ssec}

Let us derive conditions for the absence of ghosts and Laplacian 
instabilities for scalar perturbations. 
{}From Eq.~(\ref{ndef}), the perturbation of each fluid number density $n_I$, 
which is expanded up to second order, is given by 
\be
\delta n_I=\frac{\delta \rho_I}{\rho_{I,n_I}}
-\frac{({\cal N}_I \partial \chi+\partial \delta j_I)^2}
{2{\cal N}_I a^5}\,,
\ee
where $\delta \rho_I$ is the density perturbation related 
to $\delta J_I$, as
\be
\delta \rho_I=\frac{\rho_{I,n_I}}{a^3} \delta J_I\,.
\ee
The fluid sound speed squares are defined by
\be
c_I^2=\frac{n_I \rho_{I,n_I n_I}}{\rho_{I,n_I}}\,, 
\ee
which are $c_c^2=+0$, $c_b^2=+0$, and $c_r^2=1/3$ 
for CDM, baryons, and radiation, respectively.

On using the property $n_I \sqrt{-g}\,u_{I i}
=J_{I i}=J_I^0 g_{0i}+J_I^j g_{ij}={\cal N}_I 
\partial_i \chi+\partial_i \delta j_I$ for linear perturbations, 
it follows that 
\ba
\partial \delta j_{I} 
=-\mathcal{N}_{I} \left( \partial \chi + \partial v_{I} \right)\,.
\label{deljre}
\ea
This relation is used to eliminate the nondynamical variable 
$\delta j_{I}$.

In total, there are ten perturbed quantities associated with the 
scalar mode: 
$\alpha, \chi$ for the metric components, 
$\delta\phi, \psi\,(=\chi_V+\phi(t) \chi)$ for the vector field, 
and $v_{I}, \delta\rho_{I}$ (with $I=c,b,r$) for each matter component.
Expanding the action (\ref{action}) up to second order 
in scalar perturbations and integrating it by parts, the 
quadratic-order action yields
\ba
{\cal S}_{S}^{(2)} = \int {\rm d}t\, {\rm d}^{3}x
\left( L_{\rm GP} + L_{Z} + L_{M} \right)\,,
\label{SS2}
\ea
where
\ba
\hspace{-0.8cm}
L_{\rm GP} &=&{} a^{3}\, \Biggl[
 \left(w_{1}\alpha+\frac{w_{2}\delta\phi}{\phi}\right)\frac{\partial^{2}\chi}{a^{2}}
- w_{3}\,\frac{(\partial\alpha)^{2}}{a^{2}} 
+ w_{4}\alpha^{2}
-\left\{(3 H w_{1} - 2 w_{4})\frac{\delta\phi}{\phi}
- \frac{w_{3}}{a^{2} \phi}\left(\,\partial^{2} \delta\phi
+ \partial^{2}\dot{\psi} \right)
 +w_{6}\,\frac{\partial^{2}\psi}{a^{2}}\right\} \alpha 
 \nonumber \\
 \hspace{-0.8cm}
 &  & ~~~~
 -\frac{w_{3}}{4}\,\frac{(\partial\delta\phi)^{2}}{a^{2}\phi^{2}}
 +w_{5}\,\frac{(\delta\phi)^{2}}{\phi^{2}}
 -\left\{\frac{(w_{6}\phi+w_{2})\psi}{2}-\frac{w_{3}}{2}\dot{\psi}\right\}
 \frac{\partial^{2}(\delta\phi)}{a^{2}\phi^{2}}
 -\frac{w_{3}}{4\phi^{2}}\,\frac{(\partial\dot{\psi})^{2}}{a^{2}}
 +\frac{w_{7}}{2}\,\frac{(\partial\psi)^{2}}{a^{2}}
\Biggr]\,,
 \\
 \hspace{-0.8cm}
 L_{Z} &=&{}a^{3}\,\left[ \frac{\phi f_{,Z}}{\rho_c+P_c}
 \left\{ (\rho_c+P_c) \frac{\partial^{2} \chi}{a^2} 
 - \dot{\delta\rho}_{c} - 3 H \left(1 + c_{c}^{2} \right)\delta\rho_{c}\right\} v_{c}
- \phi f_{,Z} \frac{(\partial v_{c})^{2}}{2 a^{2}}
+ f_{,Z} \psi\frac{\partial^{2} \chi}{a^{2}}
+ \frac{f_{,Z}}{\rho_c+P_c} \dot{\psi} \delta\rho_{c}
\right.\nonumber \\
& &~~~~\left. 
\hspace{-0.8cm}
\qquad\,
+\frac{f_{,XZ} \phi \dot{\phi} + f_{,ZZ} \dot{\phi} + 3 f_{,Z} H}{\rho_c+P_c}
\psi \delta\rho_{c}
+ \frac{1}{2} \left(2 \phi^{3} f_{,XZ} + \phi^{2} f_{,ZZ} - \phi f_{,Z}\right)
\left(\alpha + \frac{\delta\phi}{\phi}\right)^{2}
+\frac{f_{,Z}}{2\phi a^2}(\partial\psi)^{2}
 \right] \,,\\
 \hspace{-0.8cm}
 L_{M} &=&{} a^{3}\,\sum_{I=c,b,r} \left[
\left\{ (\rho_c+P_c) \frac{\partial^{2}\chi}{a^2}-\dot{\delta\rho}_I
-3H\left(1+c_I^2 \right)\,\delta\rho_I \right\}v_I
-\frac{\rho_c+P_c}{2}\,\frac{(\partial v_I)^{2}}{a^{2}}
-\frac{c_I^2}{2(\rho_c+P_c)}(\delta \rho_I)^{2}
-\alpha \delta\rho_I \right],
\ea
with 
\ba
w_{1} &=& - \phi^{3} G_{3,X} - 2 H M_{\rm pl}^{2} \,,
\label{w1def}\\
w_{2} &=& w_{1} + 2 H M_{\rm pl}^{2} =- \phi^{3} G_{3,X}  \,, 
\label{w2def}\\
w_{3} &=& - 2 \phi^{2} q_{V} \,, 
\label{w3def}\\
w_{4} &=& \frac{1}{2} \phi^{4} f_{,XX}
- \frac{3}{2} H \phi^{3} (G_{3,X} - \phi^{2} G_{3,XX})
- 3 M_{\rm pl}^{2} H^{2} \,, 
\label{w4def}\\
w_{5} &=& w_{4} - \frac{3}{2} H (w_{1} + w_{2}) \,, 
\label{w5def}\\
w_{6} &=& \frac{1}{\phi} w_{2}=- \phi^{2} G_{3,X} \,, 
\label{w6def}\\
w_{7} &=&\frac{\dot{\phi}}{\phi^3} w_{2}
=-\dot{\phi} G_{3,X} \,.
\label{w7def}
\ea
For the variables $w_{1},\cdots, w_{7}$, the same notations 
as those given in Ref.~\cite{DeFelice:2016yws} are used. 
The contribution of intrinsic vector modes to the scalar perturbation 
equations appears only through the quantity $w_3=-2\phi^2 q_V$. 
In our theory, $q_V$ is equivalent to 1.

There are six nondynamical variables $\alpha, \chi, \delta\phi, v_{c}, v_b, v_r$,
while the dynamical perturbations correspond to the four fields 
$\psi, \delta\rho_c, \delta\rho_b, \delta\rho_r$. 
Varying the action (\ref{SS2}) with respect to the six nondynamical fields 
in Fourier space, it follows that 
\ba
\hspace{-1cm}
& &
\sum_{I=c,b,r}\delta\rho_{I} 
- 2 w_{4} \alpha 
+ \left(3 H w_{1} - 2 w_{4} \right) \frac{\delta\phi}{\phi}
+ \frac{k^{2}}{a^{2}}\left( {\cal Y}+ w_{1} \chi- w_{6} \psi 
\right) =
\left(2 \phi^{3} f_{,XZ} + \phi^{2} f_{,ZZ} - \phi f_{,Z}\right) 
\left(\alpha + \frac{\delta\phi}{\phi}\right) 
\,,
\label{eqalpha}\\
\hspace{-1cm}
& &
 \sum_{I=c,b,r} \left( \rho_I+P_I \right)v_I
+ w_{1} \alpha + w_{2} \frac{\delta\phi}{\phi} 
= - f_{,Z} \left(\phi\,v_{c} + \psi\right)\,,
\label{eqchi}\\
\hspace{-1cm}
& &
\left(3 H w_{1} - 2 w_{4}\right)\alpha -2 w_{5}\frac{\delta\phi}{\phi} 
+ \frac{k^{2}}{a^{2}} \left[
\frac{1}{2} {\cal Y}+w_2 \chi-\frac{1}{2} \left( 
\frac{w_2}{\phi}+w_6 \right)\psi
\right] =
\left(2 \phi^{3} f_{,XZ} + \phi^{2} f_{,ZZ} - \phi f_{,Z}\right) 
\left(\alpha + \frac{\delta\phi}{\phi}\right)  \,,
\label{eqdelphi}\\
\hspace{-1cm}
& &
\dot{\delta\rho_{I}}
+ 3 H \left( 1 + c_I^2 \right) \delta \rho_{I} 
+ \frac{k^{2}}{a^{2}} \left( \rho_I+P_I \right) 
\left(\chi + v_I\right) = 0 \,,\qquad 
{\rm for}~~I=c,b,r\,,
\label{eqv}
\ea
where 
\be
{\cal Y}=\frac{w_3}{\phi} \left( 
\dot{\psi}+\delta \phi+2\phi\,\alpha \right)\,.
\label{Ydef}
\ee
Variations of the action (\ref{SS2}) with respect to 
the dynamical perturbations lead to 
\ba
& &
\dot{\cal Y} + \left(H - \frac{\dot{\phi}}{\phi}\right) {\cal Y}
+ 2 \phi (w_{6} \alpha + w_{7} \psi) 
+ \left( w_{2} + w_{6} \phi \right) \frac{\delta\phi}{\phi} = 
-2f_{,Z} \left( \phi\,v_c+\psi \right)\,,
\label{eqpsi}\\
\hspace{-1cm}
& &
\dot{v}_{c} - 3 H c_{c}^{2} v_{c} 
- c_{c}^{2} \frac{\delta\rho_{c}}{\rho_{c} + P_{c}} 
- \alpha = - \frac{1}{a^{3} (\rho_{c} + P_{c})}
\frac{\partial}{\partial t} 
\left[a^{3} f_{,Z} \left( \phi\,v_{c} + \psi\right) \right] \,,
\label{eqdrhoc}\\
\hspace{-1cm}
& &
\dot{v}_{I} - 3 H c_{I}^{2} v_{I} 
- c_{I}^{2} \frac{\delta\rho_{I}}{\rho_{I} + P_{I}} 
- \alpha = 0 \,,\qquad 
{\rm for}~~I=b,r\,.
\label{eqdrhob}
\ea

We eliminate the nondynamical perturbations from the 
action (\ref{SS2}) by solving Eqs.~(\ref{eqalpha})-(\ref{eqv}) for 
$\alpha$, $\chi$, $\delta\phi$, $v_c$, $v_b$, $v_r$.
After the integration by parts, the resulting second-order action 
in Fourier space can be expressed in the form,
\be
{\cal S}_{S}^{(2)} = \int {\rm d} t\,{\rm d}^3 x\, a^{3} \left(
 \dot{\vec{\mathcal{X}}}^{t}{\bm K}
\dot{\vec{\mathcal{X}}}
-\frac{k^2}{a^2}\vec{\mathcal{X}}^{t}{\bm G}
\vec{\mathcal{X}} 
 -\vec{\mathcal{X}}^{t}{\bm M}
\vec{\mathcal{X}}
-\frac{k}{a}\vec{\mathcal{X}}^{t}{\bm B}
\dot{\vec{\mathcal{X}}}
\right) \,,
\label{SS2dynam}
\ee
where $\bm{K}$, $\bm{G}$, $\bm{M}$ and $\bm{B}$
are $4 \times 4$ matrices. 
The leading-order contributions to the matrix component 
${\bm M}$ are at most of the order $k^0$.
The vector field $\vec{\mathcal{X}}^{t}$ is composed of the dynamical 
perturbations, as
\be
\vec{\mathcal{X}}^{t}=\left( \psi,\, \delta\rho_{c}/k,\, 
\delta\rho_{b}/k,\, \delta\rho_{r}/k \right)\,.
\ee
In the small-scale limit ($k \to \infty$), the nonvanishing
components of ${\bm K}$ and ${\bm G}$ are given, 
respectively, by
\ba
& &
K_{11}=\frac{H^{2} M_{\rm pl}^{2}}{\phi^{2} (w_{1} - 2 w_{2})^{2}} 
\left[3 w_{1}^{2} + 4 M_{\rm pl}^{2} w_{4}
+ 2M_{\rm pl}^2 \left(2 \phi^{3} f_{,XZ} 
+ \phi^{2} f_{,ZZ} - \phi f_{,Z}\right) \right]
 \,,\label{K11}\\
& &
K_{22}=\frac{a^{2}(\rho_c+P_c+\phi f_{,Z})}
{2(\rho_c+P_c)^2}\,,\qquad
K_{33}=\frac{a^{2}}{2(\rho_b+P_b)} \,,\qquad
K_{44}=\frac{a^{2}}{2(\rho_r+P_r)} \,,
\ea
and 
\ba
& &G_{11}=\mathcal{G} + \dot{\mu} + H\mu
-\frac{w_2^2}{2(w_1-2w_2)^2 \phi^2} 
\sum_{I=c,b,r} \left( \rho_I+P_I \right)
- \frac{4 f_{,Z} H^{2}M_{\rm pl}^{4}}{2(w_1-2w_2)^2 \phi} \,,\\
& &G_{22}=\frac{a^{2}c_c^2}{2(\rho_c+P_c)}  \,,\qquad
G_{33}=\frac{a^{2}c_b^2}{2(\rho_b+P_b)}  \,,\qquad
G_{44}=\frac{a^{2}c_r^2}{2(\rho_r+P_r)} \,,
\ea
where
\be
\mathcal{G}= -\frac{4H^2M_{\rm pl}^4w_{2}^2}
{\phi^{2}w_{3}(w_{1}- 2 w_{2})^{2}} - \frac{\dot{\phi}}{2\phi^3}w_{2} \,,
\qquad 
\mu=\frac{HM_{\rm pl}^2w_{2}}{\phi^{2}(w_{1}- 2 w_{2})} \,.
\ee
The anti-symmetric matrix ${\bm B}$ has the leading-order 
off-diagonal components, which are given by 
\be
B_{12} = -B_{21} = - \frac{aH M_{\rm pl}^2 f_{,Z}}
{(w_{1} - 2 w_{2}) (\rho_{c} + P_{c})} \,.
\label{B12}
\ee
The diagonal components of ${\bm B}$ are lower than 
the order $k^0$.

In the following, we will consider perfect fluids obeying the weak 
energy conditions $\rho_I+P_I>0$ (with $I=c,b,r$).
In this case, the no-ghost conditions for baryons and radiation 
($K_{33}>0$ and $K_{44}>0$) are automatically satisfied.
The absence of ghosts for the dynamical perturbations 
$\psi$ and $\delta \rho_c$ requires that 
\ba
q_{S} &=& 3 w_{1}^{2} + 4 M_{\rm pl}^{2} w_{4}
+ 2M_{\rm pl}^2 \left(2 \phi^{3} f_{,XZ} 
+ \phi^{2} f_{,ZZ} - \phi f_{,Z}\right)>0\,,\label{nog1}\\
q_c &=&1+\frac{\phi f_{,Z}}{\rho_c+P_c}>0\,,\label{nog2}
\ea
respectively. 
By using Eq.~(\ref{Aeq}), one can easily confirm 
that $q_S$ given by Eq.~(\ref{nog1}) is identical to 
the quantity (\ref{qS0}) appearing in the denominators of 
background Eqs.~(\ref{dotphi}) and (\ref{dotH}). 
The $Z$ dependence in the coupling $f$ affects 
the no-ghost conditions of both the Proca field and CDM.

To avoid a strong-coupling problem for the Proca field, we need to impose at any time, for high $k$'s, that the diagonal term $K_{11}$ never vanishes or approaches zero. 
Similarly, the element $K_{22}\rho_c^2$ should satisfy the same no strong-coupling condition\footnote{We have multiplied $K_{22}$ by $\rho_c^2$, as this corresponds to the kinetic term for the density contrast $\delta_c=\delta\rho_c/\rho_c$.}. 
Other matter fields trivially satisfy the no strong-coupling condition.

The propagation of baryons and radiation is not modified by the 
matrix ${\bm B}$, so their sound speeds are 
$c_b^2=G_{33}/K_{33}$ and $c_r^2=G_{44}/K_{44}$, respectively. 
On the other hand, the off-diagonal components (\ref{B12}) 
affect the propagation of dynamical perturbations 
${\cal X}_1 \equiv \psi$ and 
${\cal X}_2 \equiv \delta \rho_{c}/k$.  
We substitute the solutions ${\cal X}_j=\tilde{{\cal X}}_j e^{i (\omega t-kx)}$ 
(with $j=1, 2$ and $\omega$ is a frequency) to their equations of motion 
following from the action (\ref{SS2dynam}).
To derive the dispersion relations in the small-scale limit, 
we pick up terms of the orders $\omega^2$, $\omega k$, and $k^2$. 
Then, we obtain
\ba
& & \omega^2 \tilde{{\cal X}}_1-\hat{c}_S^2\frac{k^2}{a^2}
\tilde{{\cal X}}_1-i \omega \frac{k}{a} \frac{B_{12}}{K_{11}} 
\tilde{{\cal X}}_2  \simeq 0\,,\label{dis1}\\
& & \omega^2 \tilde{{\cal X}}_2-\hat{c}_c^2\frac{k^2}{a^2}
\tilde{{\cal X}}_2-i \omega \frac{k}{a} \frac{B_{21}}{K_{22}} 
\tilde{{\cal X}}_1 \simeq 0\,,\label{dis2}
\ea
where 
\ba
\hat{c}_S^2 &=&
\frac{G_{11}}{K_{11}}=
\frac{\phi^2 (w_1-2w_2)^2}{H^2 M_{\rm pl}^2 q_S} \left[
\mathcal{G} + \dot{\mu} + H\mu
-\frac{w_2^2}{2(w_1-2w_2)^2 \phi^2} 
\sum_{I=c,b,r} \left( \rho_I+P_I \right)
- \frac{4 f_{,Z} H^{2}M_{\rm pl}^{4}}{2(w_1-2w_2)^2 \phi} \right]
\,,\label{css}\\ 
\hat{c}_c^2 &=&
\frac{G_{22}}{K_{22}}=\frac{c_c^2}{q_c} \,.
\label{csc}
\ea
Since we are considering the case $c_c^2=+0$, it follows that 
$\hat{c}_c^2=+0$. 
Then, the two solutions to Eq.~(\ref{dis2}) are given by 
\ba
& &
\omega=0\,,\label{branch1}\\
& &
\omega \tilde{\cal X}_2=i \frac{k}{a}
\frac{B_{21}}{K_{22}}
\tilde{{\cal X}}_1\,.\label{branch2}
\ea
The CDM has the dispersion relation (\ref{branch1}), so 
its sound speed squared $c_{\rm CDM}^2=\omega^2 a^2/k^2$ is
\be
c_{\rm CDM}^2=+0\,.
\ee
The perturbation $\psi$ associated with the longitudinal scalar mode of 
$A_{\mu}$ corresponds to the other branch (\ref{branch2}), so 
substitution of Eq.~(\ref{branch2}) into Eq.~(\ref{dis1})
results in the dispersion relation 
$\omega^2=c_S^2 k^2/a^2$, with 
\be
c_{S}^{2} = 
\hat{c}_{S}^{2} +\Delta c_S^2\,,
\label{cssT}
\ee
where 
\be
\Delta c_S^2=\frac{B_{12}^2}{K_{11}K_{22}}=
\frac{2 M_{\rm pl}^2 (\phi f_{,Z})^{2}}
{q_{S} q_{c}(\rho_c+P_c)}\,.
\label{Deltacs}
\ee
Thus the interaction between the Proca field and CDM gives 
rise to an additional contribution $\Delta c_S^2$ to the total 
sound speed squared $c_S^2$.
The small-scale Laplacian instability is absent for
\be
c_{S}^{2} \geq 0.
\label{cscon}
\ee
Under the no-ghost conditions (\ref{nog1}) and (\ref{nog2}), 
$\Delta c_S^2$ is positive.
This means that, as long as $\hat{c}_S^2$ defined 
by Eq.~(\ref{css}) is positive, the Laplacian 
instability is always absent for the perturbation $\psi$. 

In summary, there are neither ghosts nor Laplacian instabilities 
for scalar perturbations under the conditions (\ref{nog1}), (\ref{nog2}), and (\ref{cscon}). 
As long as $c_c^2=+0$, the coupling between the Proca field and 
CDM does not modify the effective CDM sound speed squared $c_{\rm CDM}^2$.

\section{Effective gravitational couplings for CDM and baryons}
\label{sec4}

To confront coupled dark energy models in GP theories with the 
observations of galaxy clusterings and weak lensing, we need to 
understand the evolution of matter density perturbations 
at low redshifts. For this purpose, we derive the effective gravitational couplings 
felt by CDM and baryon density perturbations by employing the so-called quasi-static 
approximation. The contribution of radiation to the background and perturbation 
equations of motion is ignored in the following discussion. 

We consider the case in which the equations of state and the 
sound speed squares of CDM and baryons are given by 
\be
w_c=0\,,\qquad w_b=0\,,\qquad
c_c^2=0\,,\qquad c_b^2=0\,.
\ee
We also introduce the CDM and baryon density contrasts, 
\be
\delta_c= \frac{\delta \rho_c}{\rho_c} \,,\qquad 
\delta_b= \frac{\delta \rho_b}{\rho_b} \,.
\label{deltacb}
\ee
From Eq.~(\ref{eqv}), we obtain
\be
\dot{\delta}_{I} 
= - \frac{k^{2}}{a^{2}} \left(\chi + v_{I}\right) \,,\qquad 
{\rm for}~I=c,b\,.
\label{eqdotdelta}
\ee
We can express Eqs. (\ref{eqdrhoc}) and (\ref{eqdrhob}) 
in the forms, 
\ba
&&
\dot{v}_{c} = \frac{1}{q_{c}}  
\left[ \alpha-\frac{H}{\phi} \left\{ q_c \epsilon_c
+(1-q_c) \epsilon_{\phi} \right\} \psi
+\frac{1}{\phi} (1-q_c) \dot{\psi} 
-H q_c \epsilon_c v_c \right]\,,
\label{dotvc}\\
&& 
\dot{v}_{b} = \alpha \,,
 \label{dotvb}
\ea
where 
\ba
q_c &=& 1+\frac{\phi f_{,Z}}{\rho_c}\,,\label{qc}\\
\epsilon_{c}&=&\frac{\dot{q_{c}}}{H q_c}
=\frac{(f_{,Z}+f_{,XZ}\phi^2+f_{,ZZ}\phi)\dot{\phi}+
3H \phi f_{,Z}}{H(\phi f_{,Z}+\rho_c)}\,, \\
\epsilon_{\phi}&=&\frac{\dot{\phi}}{H \phi}\,.
\ea
If there is no $Z$ dependence in $f$, we have  
$q_c=1$ and $\epsilon_{c}=0$, in which case 
$\dot{v}_c=\alpha$.

The gauge-invariant Bardeen potentials are defined by 
\be
\Psi=\alpha + \dot{\chi} \,,\qquad
\Phi=H \chi \,.
\label{bardeen}
\ee
Taking the time derivatives of Eq.~(\ref{eqdotdelta})
and using Eqs.~(\ref{dotvc})-(\ref{dotvb}), 
it follows that
\begin{align}
&\ddot{\delta}_{c} + (2 + \epsilon_{c}) H \dot{\delta}_{c}
+ \frac{k^{2}}{a^{2}} \frac{\Psi}{q_{c}} 
+ \frac{k^{2}}{a^{2}} \left[ 
\left( 1-\frac{1}{q_c} \right) \left( \frac{\dot{\Phi}}{H} 
-\epsilon_H \Phi \right)+\epsilon_c \Phi
\right]
 -\frac{k^{2}}{a^{2}} \frac{H}{\phi} 
\left[ \left( 1-\frac{1}{q_c} \right) 
 \left( \frac{\dot{\psi}}{H} 
-\epsilon_{\phi} \psi \right)+\epsilon_c \psi
\right] 
=0
 \label{delc_evo}
\,,\\
&\ddot{\delta}_{b} + 2 H \dot{\delta}_b 
+ \frac{k^{2}}{a^{2}} \Psi 
=0 \,,
 \label{delb_evo}
\end{align}
where 
\be
\epsilon_H=\frac{\dot{H}}{H^2}\,.
\ee
In contrast to Eq.~(\ref{delb_evo}) of baryon perturbations, 
the evolution of CDM density contrast is nontrivially affected by 
the $Z$ dependence in $f$ through the quantities containing 
$\Phi$, $\dot{\Phi}$, $\psi$, $\dot{\psi}$ in Eq.~(\ref{delc_evo}). 
By using the quasi-static approximation in the following, we derive 
the closed-form expressions of $\Psi$, $\Phi$, and $\psi$ to estimate 
the gravitational couplings of CDM and baryon density perturbations.

\subsection{Quasi-static approximation}

We employ the quasi-static approximation for the modes deep inside 
the horizon, under which the dominant contributions to the perturbation equations 
are the terms containing $k^2/a^2$ as well as $\delta \rho_c$, $\delta \rho_b$ and 
their time derivatives \cite{Boisseau:2000pr,Tsujikawa:2007gd,DeFelice:2011hq}.
Then, from Eqs.~(\ref{eqalpha}) and (\ref{eqdelphi}),
it follows that 
\begin{align}
\delta\rho_{c} + \delta\rho_{b} \simeq 
-\frac{k^{2}}{a^{2}}  \left(
{\cal Y} + w_{1} \chi - w_{6} \psi
\right) \,, 
\label{eqalpha_quasi}\\
{\cal Y} \simeq
\left(\frac{w_{2}}{\phi} - w_{6}\right) \psi
- 2 w_{2} \chi
\label{Yquasi} \,.
\end{align}
Substituting Eq.~(\ref{Yquasi}) into Eq.~(\ref{eqalpha_quasi}) and 
using $\delta_I$ ($I=c,b$) and $\Phi$ defined in Eqs.~(\ref{deltacb})
and (\ref{bardeen}), respectively, we obtain
\be
\rho_c \delta_c + \rho_b \delta_b \simeq
- \frac{k^{2}}{a^{2}} \left(
\frac{w_{1} - 2 w_{2}}{H} \Phi
+ \frac{w_{2}}{\phi} \psi \right)\,.
\label{quasi01}
\ee
{}From Eqs.~(\ref{Ydef}) and (\ref{Yquasi}), it follows that 
\be
\dot{\psi} \simeq
\frac{w_{2} + w_{6} \phi}{w_{3}} \psi 
- 2 \phi \left(\alpha + \frac{w_{2}}{w_{3}} 
\frac{\Phi}{H} \right)
- \delta\phi\,.
\label{dotpsi}
\ee

We differentiate Eq.~(\ref{quasi01}) with respect to $t$ and resort to 
Eqs.~(\ref{eqdotdelta}) and (\ref{dotpsi}) to remove $\dot{\delta}_c$, 
$\dot{\delta}_b$, and $\dot{\psi}$. 
The perturbation $\delta \phi$ can be eliminated by exploiting 
Eq.~(\ref{eqchi}). After this procedure the CDM velocity potential 
$v_c$ still remains, so we employ Eq.~(\ref{eqdotdelta}) to express 
it in terms of $\dot{\delta}_c$ and $\Phi$, as
\be
v_c=-\frac{a^2}{k^2} \dot{\delta}_c-\frac{\Phi}{H}\,.
\label{vcdel}
\ee
Then, we obtain 
\be
\phi^{2} (w_{1} - 2 w_{2}) w_{3} \Psi 
+\mu_{1} \Phi + \mu_{2} \psi \simeq 
\frac{a^{2}}{k^{2}} w_{3} \phi^{2} 
(q_c-1) \rho_c \dot{\delta}_c\,,
\label{quasi02}
\ee
where
\begin{align}
&\mu_{1} = 
\frac{\phi^2}{H} \left[
\left( \dot{w}_{1} - 2 \dot{w}_{2}
+ H w_{1} -\rho_{b} - q_{c} \rho_{c} 
\right) w_{3} - 2 w_{2} \left(
w_{2} + H w_{3} \right) \right]\,,\\
&\mu_{2} = 
\phi \left( w_{2}^{2} + H w_{2} w_{3}
+ \dot{w}_{2} w_{3} \right)
+ w_{2} \left( w_{6} \phi^{2} - w_{3} \dot{\phi} \right)
+ \phi\,w_{3} \rho_{c} (q_{c} - 1) \,.
\end{align}

We also substitute Eq.~(\ref{Yquasi}) and its time derivative 
into Eq.~(\ref{eqpsi}) by exploiting the relations 
(\ref{dotpsi}) and (\ref{vcdel}).
This procedure leads to 
\begin{align}
2 \phi^{2} w_{2} \Psi + \mu_{3} \Phi + \mu_{4} \psi
\simeq 
-\frac{2 a^{2}}{k^{2}}  \phi^{2} (q_{c} - 1) \rho_c \dot{\delta}_c\,,
\label{quasi03}
\end{align}
where
\begin{align}
\mu_{3} = &\frac{2 \phi}{H w_{3}}\, \mu_{2} \,,\\
\mu_{4} = 
&- \frac{1}{w_{3}} \left[
\phi^{3} (w_{6}^{2} + 2 w_{3} w_{7})
+ \phi^{2} \left( 2 w_{2} w_{6} + H w_{3} w_{6} + w_{3} \dot{w}_{6} \right)
+ \phi \left\{ w_{2}^{2} + H w_{2} w_{3} 
+ w_{3} \left( \dot{w}_{2} - \dot{\phi} w_{6}\right)\right\}
- 2 \dot{\phi} w_{2} w_{3}
\right]
\nonumber \\
 &- 2 \phi \rho_{c} (q_{c} - 1) \,.
\end{align}
Since $q_c-1=\phi f_{,Z}/\rho_c$, the $Z$ dependence in $f$ gives 
rise to the new terms containing $\dot{\delta}_c$ on 
the right-hand-sides of Eqs.~(\ref{quasi02}) and (\ref{quasi03}). 
Combining Eq.~(\ref{quasi02}) with (\ref{quasi03}) to eliminate the 
time derivative $\dot{\delta}_c$, we obtain
\be
2\phi^2 \left( w_1-w_2 \right) w_3 \Psi+
\left( 2\mu_1+\mu_3 w_3 \right) \Phi+
\left( 2\mu_2+\mu_4 w_3 \right) \psi=0\,.
\label{ani0}
\ee
On using the definitions of $w_1, \cdots, w_7$ in 
Eqs.~(\ref{w1def})-(\ref{w7def}) and the background 
Eqs.~(\ref{Eq00})-(\ref{Eq11}), the following equalities hold
\ba
& &
2\phi^2 \left( w_1-w_2 \right) w_3=
2\mu_1+\mu_3 w_3=-4H \phi^2 M_{\rm pl}^2 w_3\,,\\
& &
2\mu_2+\mu_4 w_3=0\,.
\ea
Then, Eq.~(\ref{ani0}) reduces to 
\be
\Psi=-\Phi\,,
\label{quasi04}
\ee
which shows the absence of an anisotropic stress.

It is convenient to introduce the two dimensionless variables, 
\ba
\alpha_{\rm B} &=& \frac{\phi^{3} G_{3,X}}{2 M_{\rm pl}^2 H} \,,
\label{aB}\\
\hat{\nu}_{S} &=& \frac{q_{S} \hat{c}_{S}^{2}}{4 M_{\rm pl}^{4} H^{2}} \,,
\label{nuS}
\ea
where 
\be
q_S \hat{c}_{S}^{2}=2M_{\rm pl}^2 \left[ H \phi^5 \epsilon_{\phi} 
G_{3,XX}+H \phi^3 (1+2\epsilon_{\phi}) G_{3,X}
-\rho_c (q_c-1) \right]-\phi^6 G_{3,X}^2 \left( 
1+\frac{4M_{\rm pl}^2}{w_3} \right)\,.
\label{hatqstcs}
\ee
Then, the quantities $w_1$, $w_2$, $\mu_1$, and $\mu_2$ 
appearing in Eqs.~(\ref{quasi01}) and (\ref{quasi02}) are 
expressed, respectively, as
\ba
& &
w_1=-2H M_{\rm pl}^2 \left( \alpha_{\rm B}+1 \right)\,,\qquad 
w_2=-2H M_{\rm pl}^2 \alpha_{\rm B}\,,\\
& &
\mu_1=2H \phi^2 M_{\rm pl}^2 w_3 \left( \alpha_{\rm B}^2
+\hat{\nu}_S -1 \right)\,,\qquad 
\mu_2=-2H^2 \phi M_{\rm pl}^2 w_3 \left( \alpha_{\rm B}^2
+\hat{\nu}_S  \right)\,.
\ea
On using Eq.~(\ref{quasi04}), we can solve Eqs.~(\ref{quasi01}) and 
(\ref{quasi02}) for $\Psi, \Phi, \psi$, as
\begin{align}
&\Psi = -\Phi 
\simeq - \frac{a^{2}}{2 M_{\rm pl}^{2} k^{2}} \left[
\left(1 + \frac{\alpha_{{\rm B}}^{2}}{\hat{\nu}_{S}}\right)
 \left( \rho_{c} \delta_{c}+\rho_{b} \delta_{b} \right)
+ \frac{\alpha_{\rm B}}{\hat{\nu}_{S}} 
\left(q_{c} - 1\right) \rho_{c} \frac{\dot{\delta}_{c}}{H}
\right] 
\,,\label{PSI}\\
&\psi \simeq 
\frac{a^{2}}{2 M_{\rm pl}^{2} k^{2}}  \frac{\phi}{H} \left[
\left\{1 + \frac{\alpha_{\rm B}(\alpha_{\rm B}  - 1)}{\hat{\nu}_{S}} \right\}
\left( \rho_{c} \delta_{c}+\rho_{b} \delta_{b} \right)
+ \frac{\alpha_{\rm B}  - 1}{\hat{\nu}_{S}} 
(q_{c} - 1) \rho_{c} \frac{\dot{\delta}_{c}}{H}
\right] \,.\label{psi}
\end{align}
The time derivatives of Eqs.~(\ref{PSI}) and (\ref{psi}) give rise to 
the terms containing $\ddot{\delta}_c$, which contribute to 
Eq.~(\ref{delc_evo}) of the CDM density contrast. 
After eliminating $\Psi$, $\dot{\Phi}$, 
$\Phi$, $\dot{\psi}$, and $\psi$ from Eq.~(\ref{delc_evo}), 
we obtain the second-order differential equation for $\delta_c$, as
\ba
&& 
\ddot{\delta}_{c} + 
H \frac{\hat{c}_S^2}{c_S^2}\left[2 + \epsilon_{c}
-\frac{3(q_c-1)\Omega_c}{2\hat{\nu}_S q_c} \left\{
(q_c-1) \left(1 + 2 \epsilon_{H} + \epsilon_{S} \right)
-2q_c \epsilon_c \right\} \right]\dot{\delta}_{c}
+\frac{3H \alpha_{\rm B}(q_c-1)}{2\hat{\nu}_S q_c}
\frac{\hat{c}_S^2}{c_S^2} \Omega_b \dot{\delta}_{b}  \nonumber \\
& &
- \frac{3 H^{2}}{2 G} \left(G_{cc} \Omega_{c} \delta_{c} 
+ G_{cb} \Omega_{b} \delta_{b}\right) \simeq 0 \,,
\label{deltaceqf}
\ea
where
\be
G_{cc} = G_{cb} =  
\left[ 1 + \frac{\alpha_{\rm B}^{2}}{\hat{\nu}_{S}}
+ \frac{\alpha_{\rm B}}{\hat{\nu}_{S}} \left\{
\left(q_{c} - 1\right)
\left(1 + \epsilon_{H} + \epsilon_{S} - \epsilon_{\rm B}\right)
- q_{c} \epsilon_{c}
\right\} \right]
\frac{1}{q_c}  \frac{\hat{c}_S^2}{c_S^2} G\,,
\label{Gcc}
\ee
with 
\be
\epsilon_{\rm B} \equiv \frac{\dot{\alpha}_{\rm B}}
{H \alpha_{\rm B}} \,, \qquad
\epsilon_{S} \equiv \frac{\dot{\hat{\nu}}_{S}}{H \hat{\nu}_{S}} \,.
\ee
{}From Eqs.~(\ref{cssT}) and (\ref{Deltacs}), the ratio between 
$c_S^2$ and $\hat{c}_S^2$ is
\be
\frac{c_S^2}{\hat{c}_S^2}=1+\frac{\Delta c_S^2}{\hat{c}_S^2}
=1+\frac{3(q_c-1)^2 \Omega_c}
{2\hat{\nu}_S q_c}\,.
\label{csr}
\ee
The difference $\Delta c_S^2$ between $c_S^2$ and $\hat{c}_S^2$, 
which arises from the off-diagonal components of matrix ${\bm B}$ 
in Eq.~(\ref{SS2dynam}), vanishes for $f_{,Z}=0$.

Substituting Eq.~(\ref{PSI}) into Eq.~(\ref{delb_evo}), we obtain
\be
\ddot{\delta}_{b} + 2 H \dot{\delta}_{b} 
-\frac{3 H\alpha_{\rm B} (q_c-1)}{2 \hat{\nu}_{S}}
\Omega_{c} \dot{\delta}_{c} 
- \frac{3 H^{2}}{2 G} \left(
G_{bc} \Omega_{c} \delta_{c}
+ G_{bb} \Omega_{b} \delta_{b}
\right)
\simeq 0 \,,
\label{deltabeqf}
\ee
where
\be
G_{bb} = G_{bc} 
= \left( 1 + \frac{\alpha_{\rm B}^{2}}{\hat{\nu}_{S}}\right) G\,.
\label{Gbb}
\ee
As long as $\hat{c}_S^2$ is positive with the absence of ghosts ($q_S>0$), 
the quantity $\hat{\nu}_S$ is positive. In coupled GP theories 
the Laplacian instability is absent for $c_S^2=\hat{c}_S^2+\Delta c_S^2>0$, so  
the condition $\hat{c}_S^2>0$ is not mandatory. 
To ensure the stability during the whole cosmic expansion history, 
however, we do not consider the special case where the two inequalities 
$\hat{c}_S^2<0$ and $c_S^2>0$ hold.
As long as $q_S \hat{c}_S^2>0$, the gravitational couplings $G_{bb}$ and $G_{bc}$  
of baryons are larger than the Newton constant $G$. 
This enhancement of $G_{bb}$ is 
attributed to the cubic-derivative coupling $G_3(X)$ \cite{DeFelice:2016uil}.
If there is no dependence of $Z$ in $f$, we have $q_c=1$, 
$\epsilon_c=0$, and $c_S^2=\hat{c}_S^2$, so the CDM gravitational coupling
(\ref{Gcc}) reduces to the value (\ref{Gbb}) of baryons.

In the presence of the coupling $f(Z)$, we observe in Eq.~(\ref{Gcc}) that 
$G_{cc}$ and $G_{cb}$ are multiplied by the factor $\hat{c}_S^2/(q_c c_S^2)$.
The quantity $q_c=1+\phi f_{,Z}/\rho_c$ should be close to 1 during 
the matter-dominated epoch ($\phi f_{,Z} \ll \rho_c$), but the magnitude 
of $q_c$ becomes greater than 1 after the dominance of the vector-field density 
as dark energy ($\phi f_{,Z} \gtrsim \rho_c$). 
Moreover, as long as $q_S\hat{c}_S^2>0$, the ratio 
$\hat{c}_S^2/c_S^2$ is smaller than 1.
Then, it is anticipated that the interaction $f(Z)$ may suppress the values 
of $G_{cc}$ and $G_{cb}$ at low redshifts. 
The term $\alpha_{\rm B}^2/\hat{\nu}_S$ in the square bracket of 
Eq.~(\ref{Gcc}) works to enhance the CDM gravitational coupling, 
but there are also additional terms proportional to $\alpha_{\rm B}$ 
in Eq.~(\ref{Gcc}).
We will show that the terms proportional to $\alpha_{\rm B}$, 
which arise from the mixture of couplings $G_3(X)$ and $f(Z)$, can play an important 
role to modify the values of $G_{cc}$ and $G_{cb}$ during the epoch 
of cosmic acceleration.
In Sec.~\ref{sec5}, we will consider a concrete model of coupled dark energy 
and investigate whether the realization of $G_{cc}$ and $G_{cb}$ 
smaller than $G$ is possible. 
Before doing so, we compute the values of $G_{cc}$ and $G_{bb}$ 
on the de Sitter background.

\subsection{Gravitational couplings on de Sitter background}

The background Eqs.~(\ref{Eq00})-(\ref{Aeq}) allow the existence of 
de Sitter solutions, along which $\phi$ and $H$ are constant with 
$\rho_I=0=P_I$. On this de Sitter background, we have
\be
\epsilon_{\phi}=0\,,\qquad \epsilon_{H}=0\,,\qquad
\epsilon_{\rm B}=0\,,\qquad \epsilon_S=0\,,\qquad
\epsilon_c=3\,.
\label{epds}
\ee
As the solutions approach the de Sitter fixed point, the quantity 
(\ref{qc}) behaves as $q_c \simeq \phi f_{,Z}/\rho_c \to \infty$, 
where the positivity of $q_c$ requires that $\phi f_{,Z}>0$. 
Of course, this behavior of $q_c$ does not mean the divergence 
of physical quantities. Indeed, on the de Sitter background satisfying 
Eq.~(\ref{epds}), Eq.~(\ref{Gcc}) reduces to 
\be
(G_{cc})_{\rm dS}=(G_{cb})_{\rm dS}
= -2\frac{\alpha_{\rm B}}{\hat{\nu}_S} \frac{\hat{c}_S^2}{c_S^2}G\,.
\label{GccdS0}
\ee
In the regime where $q_c \gg 1$, the terms proportional to 
$\alpha_{\rm B}$ in the square bracket of Eq.~(\ref{Gcc}) 
completely dominates over $\alpha_{\rm B}^2/\hat{\nu}_S$. 
This means that the gravitational coupling of CDM is very 
different from that of baryons around the de Sitter solution.
The quantities (\ref{nuS}) and (\ref{csr}) are given, 
respectively, by 
\ba
\hat{\nu}_S &=& \frac{1}{4 M_{\rm pl}^{4} H^{2}} 
\left[ 2H \phi^3 M_{\rm pl}^2 G_{3,X}
-\phi^6 G_{3,X}^2 \left( 1+\frac{4M_{\rm pl}^2}{w_3} 
\right)-2\phi M_{\rm pl}^2 f_{,Z} \right]\,,\label{qscsd}\\
\frac{c_S^2}{\hat{c}_S^2} &=&
1+\frac{\phi f_{,Z}}{2M_{\rm pl}^2 H^2 \hat{\nu}_S}\,.
\label{csra2}
\ea
As long as the condition $\hat{c}_S^2>0$ is satisfied in addition to 
the absence of ghosts ($q_S>0$ and $\phi f_{,Z}>0$), we have 
$\hat{\nu}_S=q_S \hat{c}_S^2/(4 M_{\rm pl}^{4} H^{2})>0$ and 
$c_S^2/\hat{c}_S^2>1$. Then, from Eq.~(\ref{GccdS0}), 
$(G_{cc})_{\rm dS}<0$ for $\alpha_{\rm B}>0$ and 
$(G_{cc})_{\rm dS}>0$ for $\alpha_{\rm B}<0$. 
Substituting Eqs.~(\ref{aB}), (\ref{qscsd}) and (\ref{csra2}) 
into Eq.~(\ref{GccdS0}), it follows that 
\be
(G_{cc})_{\rm dS}=(G_{cb})_{\rm dS}=\frac{4H M_{\rm pl}^2 w_3}
{\phi^3 G_{3,X}(4M_{\rm pl}^2+w_3)-2H M_{\rm pl}^2 w_3}G\,,
\label{GccdS}
\ee
while the baryon gravitational coupling (\ref{Gbb}) yields
\be
(G_{bb})_{\rm dS}=(G_{bc})_{\rm dS}=\left( 1+\frac{\phi^6 G_{3,X}^2}
{4 M_{\rm pl}^{4} H^{2} \hat{\nu}_S} \right)G\,,
\ee
where $\hat{\nu}_S$ is given by Eq.~(\ref{qscsd}).
One can express Eq.~(\ref{GccdS}) in terms of $q_V$ [see 
Eq.~(\ref{w3def})] and $\alpha_{\rm B}$, as
\be
(G_{cc})_{\rm dS}=(G_{cb})_{\rm dS}=\frac{2 q_{V} u^{2}}
{(\alpha_{\rm B}-1)q_V u^2-2\alpha_{\rm B}}G\,,
\label{GccdS2}
\ee
where 
\be
u=\frac{\phi}{M_{\rm pl}}\,.
\ee
In the expression (\ref{GccdS2}), $u$ should be evaluated on the de Sitter fixed point.
Our theory corresponds to $q_V=1$, but we explicitly write 
$q_V$ in Eq.~(\ref{GccdS2}) to accommodate more general 
intrinsic vector-mode Lagrangians like 
${\cal L}_F=-q_V F_{\mu \nu}F^{\mu \nu}/4$.
As we already mentioned, the sign of $(G_{cc})_{\rm dS}$ depends on 
$\alpha_{\rm B}$. When $\alpha_{\rm B}=1$, for example, 
we have $(G_{cc})_{\rm dS}=-q_V u^2 G$, while, for $\alpha_{\rm B} \gg 1$ 
and $q_V u^2 \gg 1$, 
$(G_{cc})_{\rm dS} \simeq (2/\alpha_{\rm B})G$. 
The self-accelerating solution in cubic-order extended Galileon scalar-tensor
theory \cite{DeFelice:2011bh,DeFelice:2011aa} can be regarded as the 
weak-coupling limit $q_V \to \infty$ 
in Eq.~(\ref{GccdS2}), so that $(G_{cc})_{\rm dS}=2G/(\alpha_{\rm B}-1)$. 
Since our coupled GP theory gives the value 
$(G_{cc})_{\rm dS}=2u^2 G/[(\alpha_{\rm B}-1)u^2-2\alpha_{\rm B}]$, 
its observational signatures associated with the cosmic growth 
measurements are different from those in its scalar-tensor counterpart.

\section{Concrete models}
\label{sec5}

To study the cosmological dynamics relevant to the late-time cosmic acceleration, 
we consider a concrete model of coupled dark energy given by 
the action (\ref{action}) with 
\be
f(X, Z) = b_{2} X^{p_{2}} + \beta (2 X)^{n}\, Z^{m} \,,
\qquad G_{3}(X) = b_{3} X^{p_{3}} \,,
\label{model}
\ee
where $b_2, b_3, p_2, p_3$ and $\beta, n, m$ 
are constants. In this model, the background Eq.~(\ref{Aeq}) yields
\be
2^{1-p_2} b_2 p_2 \phi^{2p_2-1}
+3 \cdot 2^{1-p_3} b_3 p_3 H \phi^{2p_3}
+\beta \left( 2n+m \right) \phi^{2n+m-1}=0\,.
\label{backA0}
\ee

In uncoupled GP theories ($\beta=0$), Eq.~(\ref{backA0}) shows that 
$H$ is related to $\phi$ according to 
\be
\phi^p H=\lambda={\rm constant}\,,
\label{phiH}
\ee
where $p=2p_3-2p_2+1$. 
Provided that $p>0$, the temporal vector component $\phi$ 
grows with the decrease of $H$. 
As the vector-field density dominates over the background fluid 
density,  the solutions enter the epoch of cosmic acceleration 
and finally approach the de Sitter fixed point characterized by 
constant $\phi$ \cite{DeFelice:2016yws}.

In coupled GP theories which contain the $Z$ dependence in $f$, we 
would like to consider the cosmological background possessing the 
same property as Eq.~(\ref{phiH}). 
This can be realized for the powers,
\be
p_3 = \frac{1}{2} \left(p + 2 p_{2} -1\right) \,,\qquad
n=p_2-\frac{m}{2}\,.
\label{power}
\ee
In this case, the three terms in Eq.~(\ref{backA0}) have the same 
power-law dependence of $\phi$. 
Then, from Eq.~(\ref{backA0}), the constants $b_2$, $b_3$, and 
$\beta$ are related with each other, as
\be
b_3=-\frac{2^{(p+1)/2}p_2 (b_2+2^{p_2}\beta)}
{3\lambda (p+2p_2-1)}\,.
\ee
In the following, we study the dynamics of background and perturbations 
for the functions (\ref{model}) with the powers (\ref{power}). 

\subsection{Background dynamics and theoretically consistent conditions}

To study the background dynamics, we take CDM, baryons, and 
radiation into account as perfect fluids.
The dark energy density parameter defined in 
Eq.~(\ref{Omega}) yields
\be
\Omega_{\rm DE} 
= -\frac{(2^{-p_{2}}b_{2} + \beta)\phi^{2p_{2}}}
{3 M_{\rm pl}^{2} H^{2}}  \,.
\label{OmegaDE}
\ee
By imposing the condition $\Omega_{\rm DE}>0$, 
the constants $b_2$ and $\beta$ are constrained to be 
\be
2^{-p_{2}}b_{2} + \beta<0\,.
\label{b2b}
\ee
{}From Eq.~(\ref{tOmega}), we have
\be
\Omega_b=1-\Omega_{\rm DE}-\Omega_c-\Omega_r\,.
\label{Omebcon}
\ee
On using Eqs.~(\ref{dotphi}) and (\ref{dotH}), it follows that 
\ba
\epsilon_{\phi} &=& \frac{3-3\Omega_{\rm DE}+\Omega_r}
{2p (1+s \Omega_{\rm DE})}\,,\\
\epsilon_{H} &=& -\frac{3-3\Omega_{\rm DE}+\Omega_r}
{2(1+s \Omega_{\rm DE})}\,,
\ea
where 
\be
s=\frac{p_2}{p}\,.
\ee
Then, the density parameters $\Omega_{\rm DE}$, $\Omega_{c}$, 
and $\Omega_{r}$ obey the differential equations, 
\begin{align}
\Omega'_{\rm DE}
&= \frac{(1 + s)\,\Omega_{\rm DE}
(3 - 3 \Omega_{\rm DE}+ \Omega_{r})}{1 + s\, \Omega_{\rm DE}}\,,
\label{autoDE}\\
\Omega'_{c} 
&= \frac{\Omega_{c}\left[\Omega_{r} 
-3 (1 + s) \Omega_{\rm DE}\right]}{1 + s\,\Omega_{\rm DE}}\,,
\label{autob}\\
\Omega'_{r} 
&= - \frac{\Omega_{r} \left[1 - \Omega_{r} 
+ (3 + 4 s) \Omega_{\rm DE}\right]}{1 + s\, \Omega_{\rm DE}} \,,
\label{autor}
\end{align}
where a prime represents a derivative with respect to ${\cal N}=\ln a$. 
For a given value of $s$ and initial conditions of $\Omega_{\rm DE}$, $\Omega_c$, 
and $\Omega_r$, each density parameter is known by integrating
Eqs.~(\ref{autoDE})-(\ref{autor}) with Eq.~(\ref{Omebcon}).

The dark energy equation of state in Eq.~(\ref{wde}) and effective equation of 
state in Eq.~(\ref{weff}) are given by 
\ba
w_{\rm DE}
&=& -\frac{3(1+s)+s\Omega_r}{3(1+s\Omega_{\rm DE})}\,,
\label{wde2}\\
w_{\rm eff}
&=& \frac{\Omega_r-3(1+s) \Omega_{\rm DE}}
{3(1+s\Omega_{\rm DE})}\,,
\ea
respectively.
Apart from the fact that nonrelativistic matter is separated into 
CDM and baryons, the background dynamics is the same as that 
studied in Ref.~\cite{DeFelice:2016yws}.
As we observe in Eq.~(\ref{OmegaDE}), the effect of new coupling 
$\beta$ can be simply absorbed into the definition of $\Omega_{\rm DE}$ 
at the background level. 

\begin{figure}[ht]
\begin{center}
\includegraphics[height=3.4in,width=3.4in]{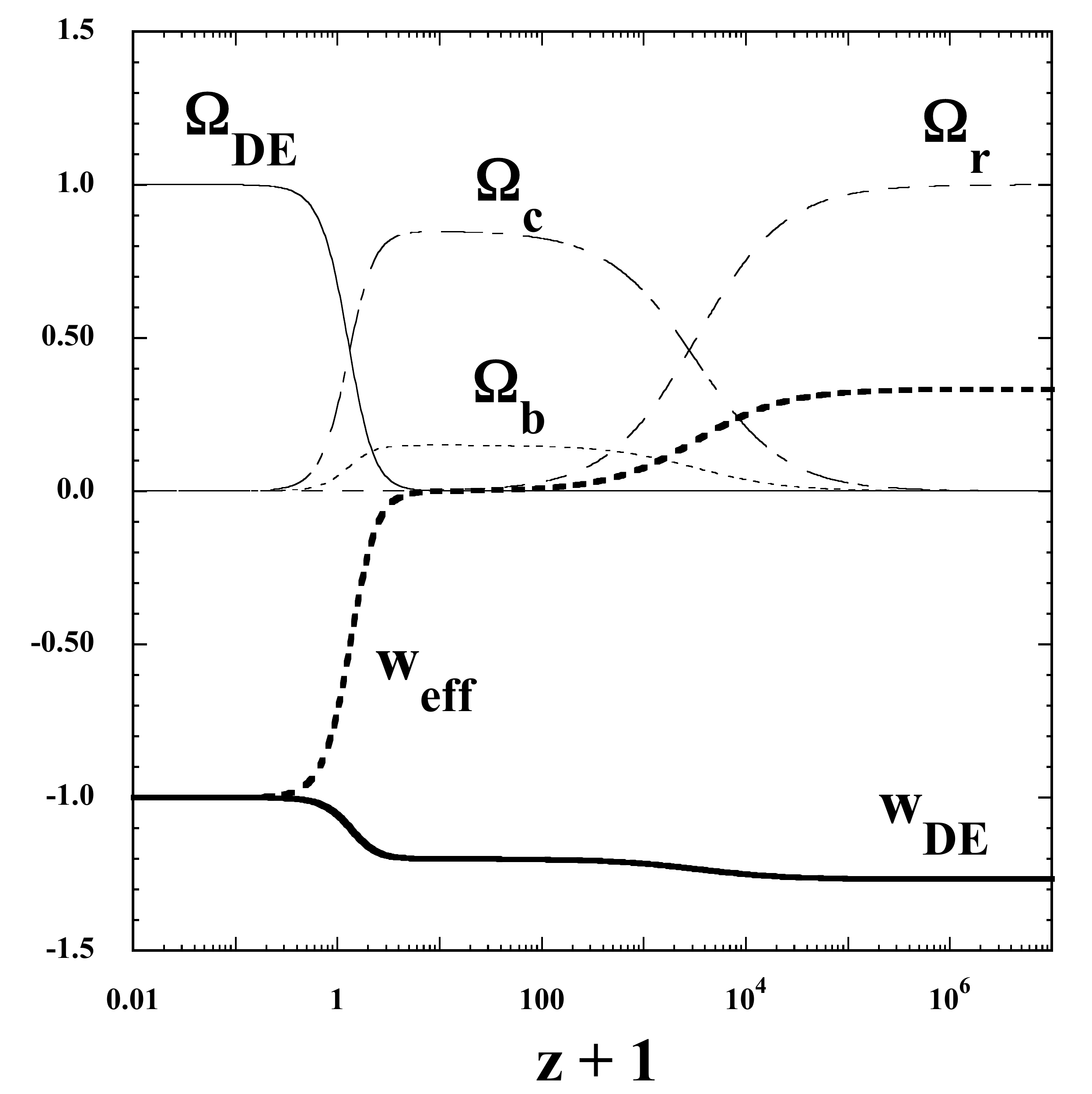}
\includegraphics[height=3.4in,width=3.5in]{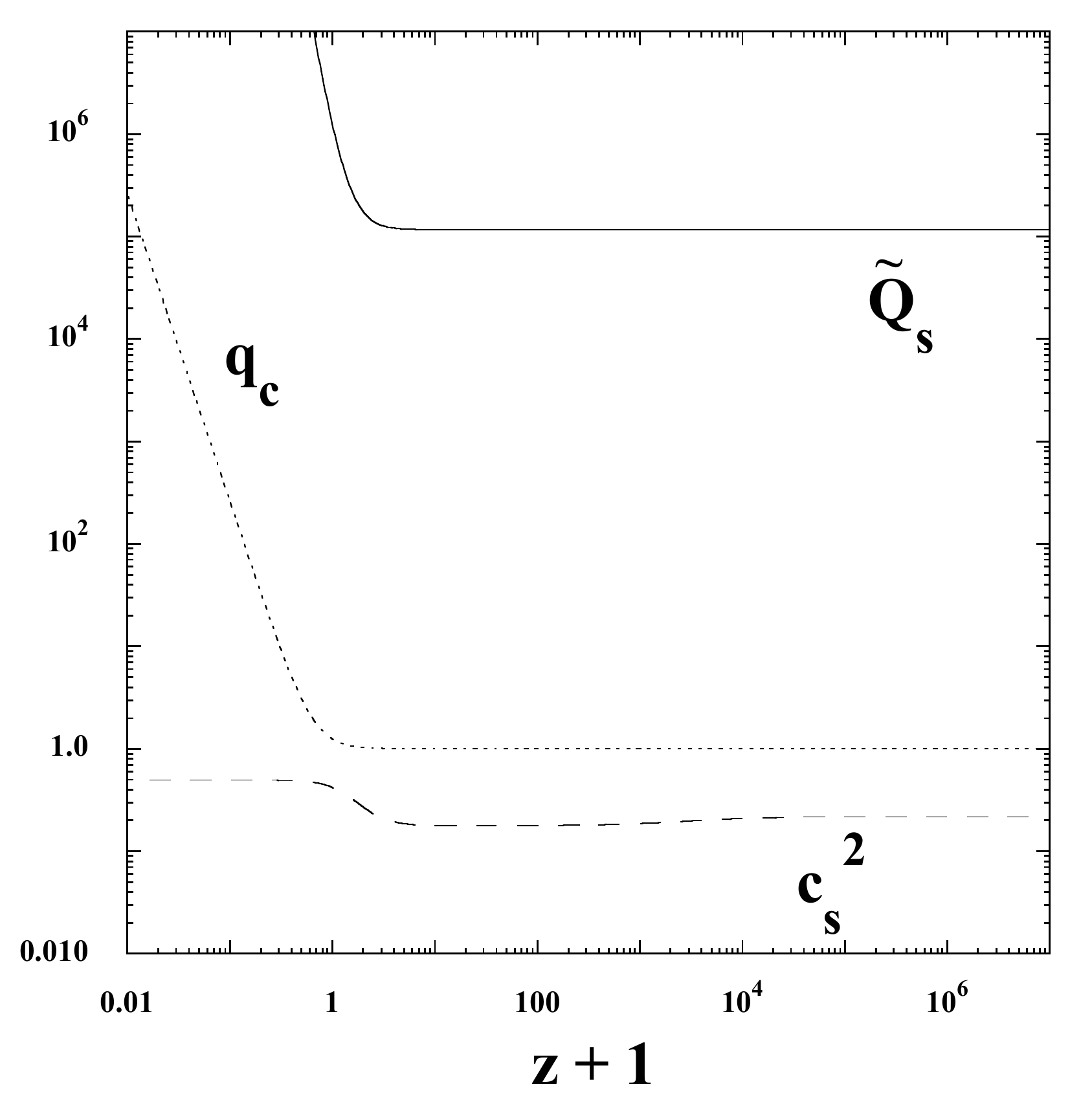}
\end{center}
\caption{(Left) Evolution of $w_{\rm DE}$, $w_{\rm eff}$ and 
$\Omega_{\rm DE}$, $\Omega_{c}$, $\Omega_b$, $\Omega_r$ 
versus $z+1$ for $s=1/5$, where $z=1/a-1$ is the redshift 
with today's scale factor $a=1$.
The initial conditions of $\Omega_{\rm DE}$, $\Omega_{c}$, 
and $\Omega_r$ are chosen to realize their today's values 
$\Omega_{\rm DE}(z=0)=0.68$, $\Omega_c(z=0)=0.27$, 
$\Omega_b(z=0)=0.05$, and $\Omega_r(z=0)=10^{-4}$, respectively. 
(Right) Evolution of $q_c$, $\tilde{Q}_S=K_{11} M_{\rm pl}^{2p}/\lambda^2$, 
and $c_S^2$ for $p_2=1$, $p=5$, $m=2$, and $r_{\beta}=0.05$ 
with the same initial conditions of density parameters as those 
used in the left panel, with today's dimensionless temporal 
vector component $u(z=0)=0.459$.
\label{fig1}}
\end{figure}

During the cosmological sequence of radiation ($\Omega_r=1$, $w_{\rm eff}=1/3$), 
matter ($\Omega_c+\Omega_b=1$, $w_{\rm eff}=0$), and de Sitter 
($\Omega_{\rm DE}=1$, $w_{\rm eff}=-1$) epochs, 
the dark energy equation of state (\ref{wde2}) changes as 
$w_{\rm DE}=-1-4s/3 \to -1-s \to -1$, respectively, see 
the left panel of Fig.~\ref{fig1} for the case $s=1/5$.
Thus the background dynamics is solely determined by the 
single parameter $s$, which characterizes the deviation from 
the $\Lambda$CDM model.

We define the density parameter associated with 
the coupling $\beta$, as
\be
\Omega_{\beta}=\frac{\beta \phi^{2p_2}}
{3 M_{\rm pl}^2 H^2}\,.
\ee
Then, the no-ghost conditions (\ref{nog1}) and (\ref{nog2}) 
translate, respectively, to 
\ba
q_S &=& 12 M_{\rm pl}^4 H^2 p^2 s \Omega_{\rm DE} 
\left( 1+s \Omega_{\rm DE} \right)>0\,,\label{qs1} \\
q_c &=& 1+\frac{m\Omega_{\beta}}{\Omega_c}>0\,.
\label{qc1}
\ea
To satisfy the condition (\ref{qs1}) in the asymptotic 
past ($\Omega_{\rm DE} \to +0$), the parameter $s$ 
is in the range, 
\be
s > 0\,.
\label{scon}
\ee
This means that $w_{\rm DE}$ is always in the phantom 
region ($w_{\rm DE}<-1$). 
Around the future de Sitter fixed point, the parameter (\ref{qc1}) 
behaves as $q_c \simeq m\Omega_{\beta}/\Omega_c$, so 
its positivity requires that 
\be
m\Omega_{\beta}>0\,.
\label{mO}
\ee
For positive $m$, the inequality (\ref{mO}) implies that $\beta>0$.
The condition (\ref{mO}) is not obligatory for the cosmic 
expansion history by today, but we impose it to ensure the
stability around the future de Sitter solution.

As for the no strong-coupling condition, the quantity 
given by Eq.~(\ref{K11}) reduces to
\be
K_{11}=\frac{3p^{2}s\,M_{\rm pl}^2 H^{2} \Omega_{{\rm DE}}
(1+s\Omega_{{\rm DE}})}{(1-ps\Omega_{{\rm DE}})^{2}\,\phi^{2}}\,.
\label{K11d}
\ee
At early times ($\Omega_{\rm DE} \ll 1$), $K_{11}$ has the dependence, 
\be
K_{11} \propto \Omega_{\rm DE}^{(ps-1)/[p(s+1)]}\,,
\label{K11d2}
\ee
so that the strong coupling can be avoided for 
\be
0<p\,s\leq1\,,\qquad{\rm or}\qquad 0<p_2\leq1\,.
\ee
We remind the reader that we are considering the case $p>0$, 
in order for the Proca field to be responsible for the late-time cosmic acceleration.

During the radiation, matter, and de Sitter epochs, the sound 
speed squared (\ref{cssT}) reduces, respectively, to 
\ba
(c_{S}^{2})_{\rm ra} &=& 
\frac{p(3 + 4 s) - 2}{3 p^{2}}-\frac{m r_{\beta}}{2p^2 s}\,,
\label{csrad}\\
(c_{S}^{2})_{\rm ma} &=&
\frac{p(5+6s) - 3}{6 p^{2}}-\frac{m r_{\beta}}{2p^2 s}\,,
\label{csma}\\
(c_{S}^{2})_{\rm dS} &=& 
\frac{1}{3p(1+s)} \left( 1-ps -\frac{4ps M_{\rm pl}^2}{w_3} 
\right)\,,\label{csds}
\ea
where 
\be
r_{\beta}=\frac{\Omega_{\beta}}{\Omega_{\rm DE}}
=-\frac{\beta}{2^{-p_2}b_2+\beta}\,.
\ee
As long as $\Omega_{\rm DE}>0$, the condition (\ref{mO}) 
translates to $m r_{\beta}>0$.
The constant $r_{\beta}$ characterizes the contribution of
the coupling $\beta$ to the total dark energy density.
We note that the difference (\ref{Deltacs}) between $c_S^2$ 
and $\hat{c}_S^2$ is given by 
\be
\Delta c_S^2=\frac{m^2 r_{\beta}^2\Omega_{\rm DE}}{2p^2 s
(\Omega_c+m r_{\beta} \Omega_{\rm DE})
(1+s \Omega_{\rm DE})}\,.
\label{Delces}
\ee
This quantity vanishes on the radiation and matter fixed points 
($\Omega_{\rm DE}=0$), so $(\hat{c}_{S}^{2})_{\rm ra}$ and 
$(\hat{c}_{S}^{2})_{\rm ma}$ are identical to 
$(c_{S}^{2})_{\rm ra}$ and $(c_{S}^{2})_{\rm ma}$, respectively. 
On the de Sitter solution, there is the difference 
$(\Delta c_S^2)_{\rm dS}=m r_{\beta}/[2p^2s (1+s)]$, so that 
\be
(\hat{c}_{S}^{2})_{\rm dS} = 
\frac{1}{3p(1+s)} \left( 1-ps -\frac{4ps M_{\rm pl}^2}{w_3} 
\right)
-\frac{m r_{\beta}}{2p^2s (1+s)}\,.
\label{hatcs2}
\ee
In Eq.~(\ref{csds}), the coupling $\beta$ disappears from 
$(c_S^2)_{\rm dS}$ due to the contribution 
$(\Delta c_S^2)_{\rm dS}$ to $(\hat{c}_{S}^{2})_{\rm dS}$. 
To avoid the Laplacian instability during 
the whole cosmological evolution, we require that $(c_{S}^{2})_{\rm ra}$, 
$(c_{S}^{2})_{\rm ma}$, and $(c_{S}^{2})_{\rm dS}$ are all positive.

In the right panel of Fig.~\ref{fig1}, we plot the evolution of 
$q_c$, $\tilde{Q}_S=K_{11} M_{\rm pl}^{2p}/\lambda^2$, and $c_S^2$ 
for the model parameters $p_2=1$, $p=5$, $m=2$, and 
$r_{\beta}=0.05$. Today's values of density parameters 
(at the redshift $z=0$)
are the same as those in the left panel, with 
$u(z=0)=\phi (z=0)/M_{\rm pl}=0.459$. 
Since $s~(=1/5)$, $m$, $\Omega_{\rm DE}$, and 
$r_{\beta}=\Omega_{\beta}/\Omega_{\rm DE}$ are all positive, 
the no-ghost conditions (\ref{qs1}) and (\ref{qc1}) are 
automatically satisfied. 
Indeed, the positivities of $\tilde{Q}_S$ and $q_c$ can be confirmed 
in Fig.~\ref{fig1}.
Since the numerical simulation of Fig.~\ref{fig1} corresponds to 
$ps=1$, $K_{11}$ stays constant 
in the asymptotic past ($\Omega_{\rm DE} \ll 1$), see Eq.~(\ref{K11d2}). 
As we observe in Fig.~\ref{fig1}, the quantity 
$\tilde{Q}_S=K_{11} M_{\rm pl}^{2p}/\lambda^2$ 
continues to grow toward 
the future de Sitter attractor, so there is no strong-coupling problem 
for the Proca field.  This is also the case for CDM, where 
the quantity $K_{22}\rho_c^2=a^2 (\rho_c+\phi f_{,Z})/2$ 
approaches 0 neither in the asymptotic past nor in the future.

For the model parameters used in the numerical simulation of 
Fig.~\ref{fig1}, the analytic estimations (\ref{csrad}) and (\ref{csma}) 
give  $(c_{S}^{2})_{\rm ra}=0.217$ and 
$(c_{S}^{2})_{\rm ma}=0.177$, which agree well with
their numerical values in Fig.~\ref{fig1}.
On using the asymptotic value $u_{\rm dS}=\phi_{\rm dS}/M_{\rm pl}=0.474$ on 
the de Sitter solution, we obtain $(c_{S}^{2})_{\rm dS}=0.494$ and 
$(\hat{c}_S^2)_{\rm dS}=0.485$ 
from Eqs.~(\ref{csds}) and (\ref{hatcs2}). 
Again, they are in good agreement with their numerical values. 
As we observe in Fig.~\ref{fig1}, the scalar sound speed squared 
$c_S^2$ is always positive from the radiation era to the de Sitter epoch. 
Hence, for the model parameters and initial conditions 
used in Fig.~\ref{fig1}, we realize a viable cosmology 
without ghosts or Laplacian instabilities.

\subsection{Dynamics of matter perturbations}

We proceed to the study of matter density perturbations relevant to the 
observations of galaxy clusterings, weak lensing, and CMB.
Since we are interested in the late-time evolution of perturbations, 
we ignore the contributions of radiation to the background and 
perturbation equations.

During the matter-dominated epoch in which $\Omega_{\rm DE}$ is 
less than the order 1, we compute the CDM and baryon gravitational couplings 
by expanding Eqs.~(\ref{Gcc}) and (\ref{Gbb}) in terms of $\Omega_{\rm DE}$. 
Then, it follows that 
\ba
& &
(G_{cc})_{\rm ma}=(G_{cb})_{\rm ma}=\left[ 1+
{\cal F}\Omega_{\rm DE}
+{\cal O} \left( \Omega_{\rm DE}^2 \right)\right] G\,, 
\label{Gccea}\\
& &
(G_{bb})_{\rm ma}=(G_{bc})_{\rm ma}=\left[ 
1+ \frac{s}{3 (c_S^2)_{\rm ma}} \Omega_{\rm DE} 
+{\cal O} \left( \Omega_{\rm DE}^2 \right)
\right] G\,,
\label{Gbbea}
\ea
where 
\be
{\cal F}=
\frac{s}{3 (c_S^2)_{\rm ma}} -\frac{m r_{\beta}\{ 4p(1+s)-1\}}
{2p^2 (c_S^2)_{\rm ma} \Omega_c}\,,
\label{calF}
\ee
and $(c_S^2)_{\rm ma}$ is given by Eq.~(\ref{csma}).
In the early matter era ($\Omega_{\rm DE} \ll 1$), 
both $(G_{cc})_{\rm ma}$ and $(G_{bb})_{\rm ma}$ are close to $G$.
With the increase of $\Omega_{\rm DE}$, the gravitational couplings 
(\ref{Gccea}) and (\ref{Gbbea}) start to deviate from $G$. 
Since the factor $s/[(3c_S^2)_{\rm ma}]$ in Eq.~(\ref{Gbbea}) is positive under 
the absence of ghosts and Laplacian instabilities, $(G_{bb})_{\rm ma}$ 
is larger than $G$.
 
For $(G_{cc})_{\rm ma}$ given in Eq.~(\ref{Gccea}), there is an extra term 
arising from the coupling $\beta$ besides the positive factor 
$s/[(3c_S^2)_{\rm ma}]$. 
As long as $m r_{\beta}\{ 4p(1+s)-1\}>0$, the coupling $\beta$ works 
to reduce $(G_{cc})_{\rm ma}$. 
If ${\cal F}<0$ in the early matter era ($\Omega_c \simeq 1$), 
the factor ${\cal F}$ remains negative due to the decrease of $\Omega_c$. 
If ${\cal F}>0$ initially, then there is the moment at which ${\cal F}$ 
crosses 0. This moment of transition can be quantified by 
the CDM density parameter, as 
\be
\Omega_c^{\rm T}=\frac{3m r_{\beta} 
[4p(1+s)-1]}{2p^2 s}\,.
\label{OmecT}
\ee
After $\Omega_c$ drops below $\Omega_c^{\rm T}$, 
$G_{cc}$ becomes smaller than $G$.
This transition from $G_{cc}>G$ 
to $G_{cc}<G$ occurs for the model parameters 
satisfying $\Omega_c^{\rm T}<1$, i.e., 
$2p^2 s> 3m r_{\beta} [4p(1+s)-1]$. 
We note that, if $\Omega_c^{\rm T}$ is much smaller than 1, 
the expansion of $G_{cc}$ of Eq.~(\ref{Gccea}) up to first order 
in $\Omega_{\rm DE}$ loses its validity. We are interested in the case where 
the weak gravitational interaction for CDM ($G_{cc}<G$) is realized 
by today. In this case, $\Omega_c^{\rm T}$ 
is larger than today's CDM density 
parameter $\Omega_c(z=0) \simeq 0.27$, so that 
\be
\Omega_c^{\rm T}>0.27\,,
\label{Omeccon}
\ee
which can be regarded as a criterion for the realization of weak gravity.

The parameter $\alpha_{\rm B}$ defined in Eq.~(\ref{aB}) is related to 
$\Omega_{\rm DE}$, as 
\be
\alpha_{\rm B} = p_{2} \Omega_{\rm DE}\,.
\ee
Since we are considering the theory with $q_V=1$, the CDM 
gravitational coupling (\ref{GccdS2}) on the de Sitter background 
reduces to 
\be
(G_{cc})_{\rm dS}=(G_{cb})_{\rm dS}=
\frac{2u_{\rm dS}^2}{(p_2-1)u_{\rm dS}^2-2p_2} G\,,
\label{Gccdsf}
\ee
where $u_{\rm dS}=\phi_{\rm dS}/M_{\rm pl}$.
Meanwhile, the baryon gravitational coupling (\ref{Gbb})
on the de Sitter solution yields
\be
(G_{bb})_{\rm dS}=(G_{bc})_{\rm dS}=
\left[ 1+\frac{s}{3(1+s) (\hat{c}_S)^2_{\rm dS}} \right]G\,,
\label{Gbbdsf}
\ee
where $(\hat{c}_S)^2_{\rm dS}$ is given by Eq.~(\ref{hatcs2}).
As expected,  $(G_{bb})_{\rm dS}$ is always larger than $G$, but 
this is not the case for $(G_{cc})_{\rm dS}$.

\begin{figure}[ht]
\begin{center}
\includegraphics[height=3.4in,width=3.4in]{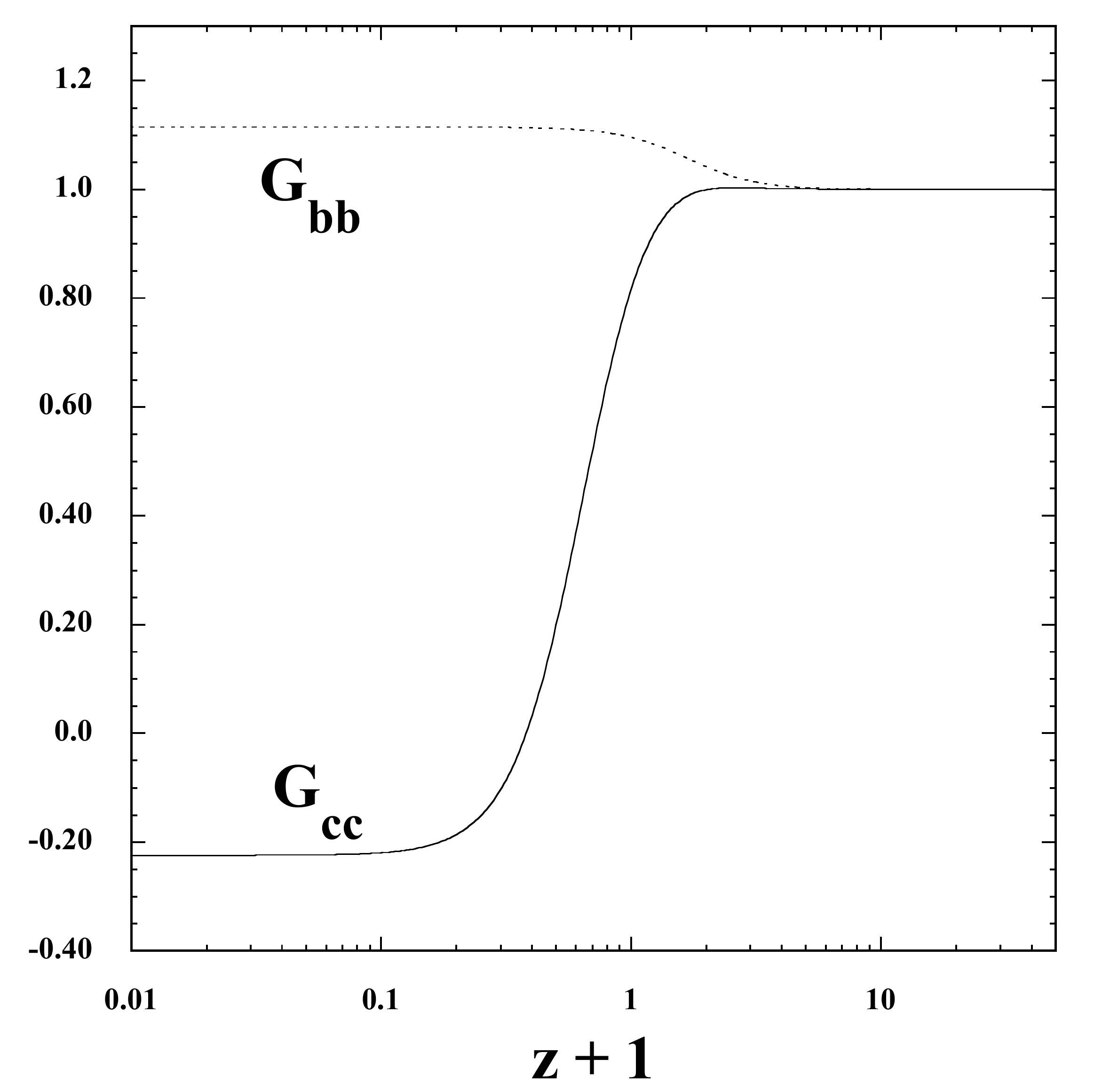}
\includegraphics[height=3.4in,width=3.5in]{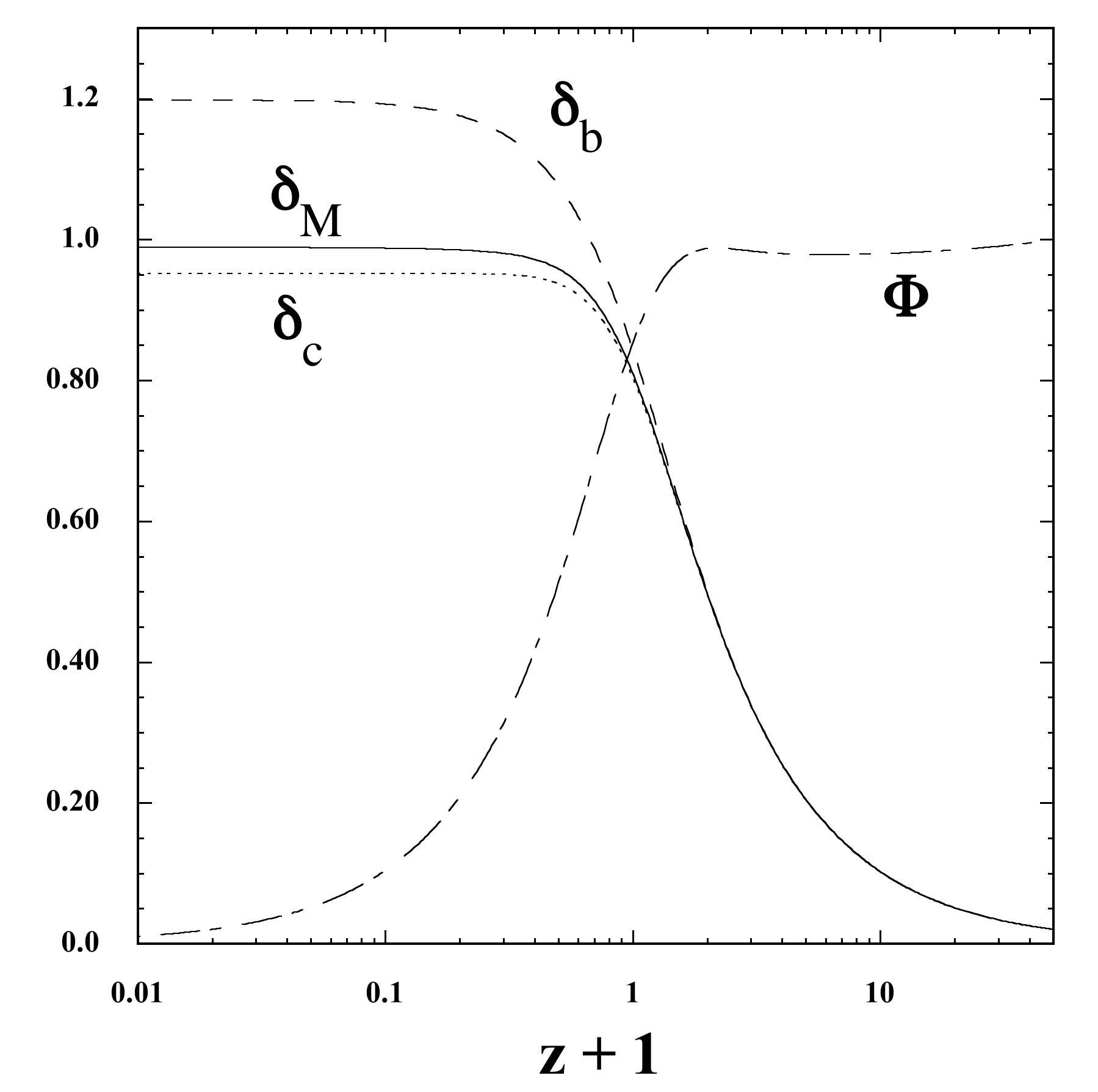}
\end{center}
\caption{Evolution of $G_{cc}$, $G_{bb}$ (left) and 
$\delta_c$, $\delta_b$, $\delta_M$, $\Phi$ (right)
versus $z+1$ for $p_2=1$, $s=1/5$, $m=2$, and 
$r_{\beta}=0.05$, with the same initial 
conditions of density parameters 
as those used in Fig.~\ref{fig1}. 
We choose today's value of the total matter density contrast 
$\delta_M$, as $\sigma_8(z=0)=0.811$. 
The gravitational potential $\Phi$ is normalized by its 
initial value at $z=50$.
\label{fig2}}
\end{figure}

In the left panel of Fig.~\ref{fig2}, we show the evolution of 
$G_{cc}$ and $G_{bb}$ for $z<50$ by using the same model 
parameters and initial conditions as those given in the 
caption of Fig.~\ref{fig1}. 
At high redshifts, we have $\Omega_{\rm DE} \ll 1$ and 
hence both $G_{cc}$ and $G_{bb}$ are close to $G$ from 
Eqs.~(\ref{Gccea}) and (\ref{Gbbea}). 
In this case the quantity (\ref{calF}) is given by 
${\cal F}=0.377-0.260/\Omega_c$, so ${\cal F}$ is 
initially positive. The CDM density parameter (\ref{OmecT}) 
at which ${\cal F}$ crosses 0 is $\Omega_c^{\rm T}=0.69$. 
Numerically, we find that $G_{cc}$ becomes smaller than $G$ 
at the redshift $z<1.06$. The numerical value of CDM 
density parameter at $z=1.06$ is $\Omega_c=0.71$, 
which is close to $\Omega_c^{\rm T}=0.69$ derived by 
the analytic estimation (\ref{OmecT}). 
As we observe in Fig.~\ref{fig2}, 
$G_{cc}$ starts to be smaller than $G$ at $z=1.06$ and 
decreases toward an asymptotic negative constant after 
crossing $G_{cc}=0$.
Since this case corresponds to $p_2=1$ in Eq.~(\ref{Gccdsf}),   
we have $(G_{cc})_{\rm dS}=-u_{\rm dS}^2 G=-0.225G$, 
where we used the numerical value $u_{\rm dS}=0.4743$ 
on the de Sitter attractor.
This analytic estimation of $(G_{cc})_{\rm dS}$ is 
in good agreement with the asymptotic numerical value 
seen in Fig.~\ref{fig2}. 
As we estimated in Eqs.~(\ref{Gbbea}) and (\ref{Gbbdsf}), 
the baryon gravitational coupling $G_{bb}$ is always 
larger than $G$. For the model parameters used in Fig.~\ref{fig2}, 
we have $(G_{bb})_{\rm dS}=1.114G$ from Eq.~(\ref{Gbbdsf}), 
which agrees well with the numerical result.

For larger $m r_{\beta}$, the density parameter (\ref{OmecT}) 
at transition tends to be larger, so that the CDM perturbation enters 
the regime $G_{cc}<G$ earlier. This means that, for increasing values 
of $m$ and $\beta$, the realization of weak gravity by the momentum transfer 
starts to occur from higher redshifts.
The gravitational coupling (\ref{Gccdsf}) on the de Sitter background
depends on $p_2$ and $u_{\rm dS}$. 
Meanwhile, the condition for the no strong-coupling problem at early times imposes that  
$0<p_2\leq1$, under which the denominator of Eq.~(\ref{Gccdsf}) is always negative. 
Then, $(G_{cc})_{\rm dS}$ is negative, as seen in 
the numerical simulation of Fig.~\ref{fig2}.
In this case the gravitational interaction is no longer attractive, 
by reflecting the fact that CDM interacts with the self-accelerating 
vector field through the momentum transfer. 
As we mentioned in Sec.~\ref{sec4}, this behavior of 
$(G_{cc})_{\rm dS}$ is mostly attributed to the mixture of 
couplings $G_3(X)$ and $f(Z)$, i.e., the terms proportional to 
$\alpha_{\rm B}$ in Eq.~(\ref{Gcc}). 
Today's CDM gravitational coupling depends on when the transition to 
the regime $G_{cc}<G$ occurs as well as on the value of $(G_{cc})_{\rm dS}$.
The numerical simulation of Fig.~\ref{fig2} corresponds to
$G_{cc}(z=0)=0.815G$, with $G_{bb}(z=0)=1.095G$.

\begin{figure}[ht]
\begin{center}
\includegraphics[height=3.3in,width=3.4in]{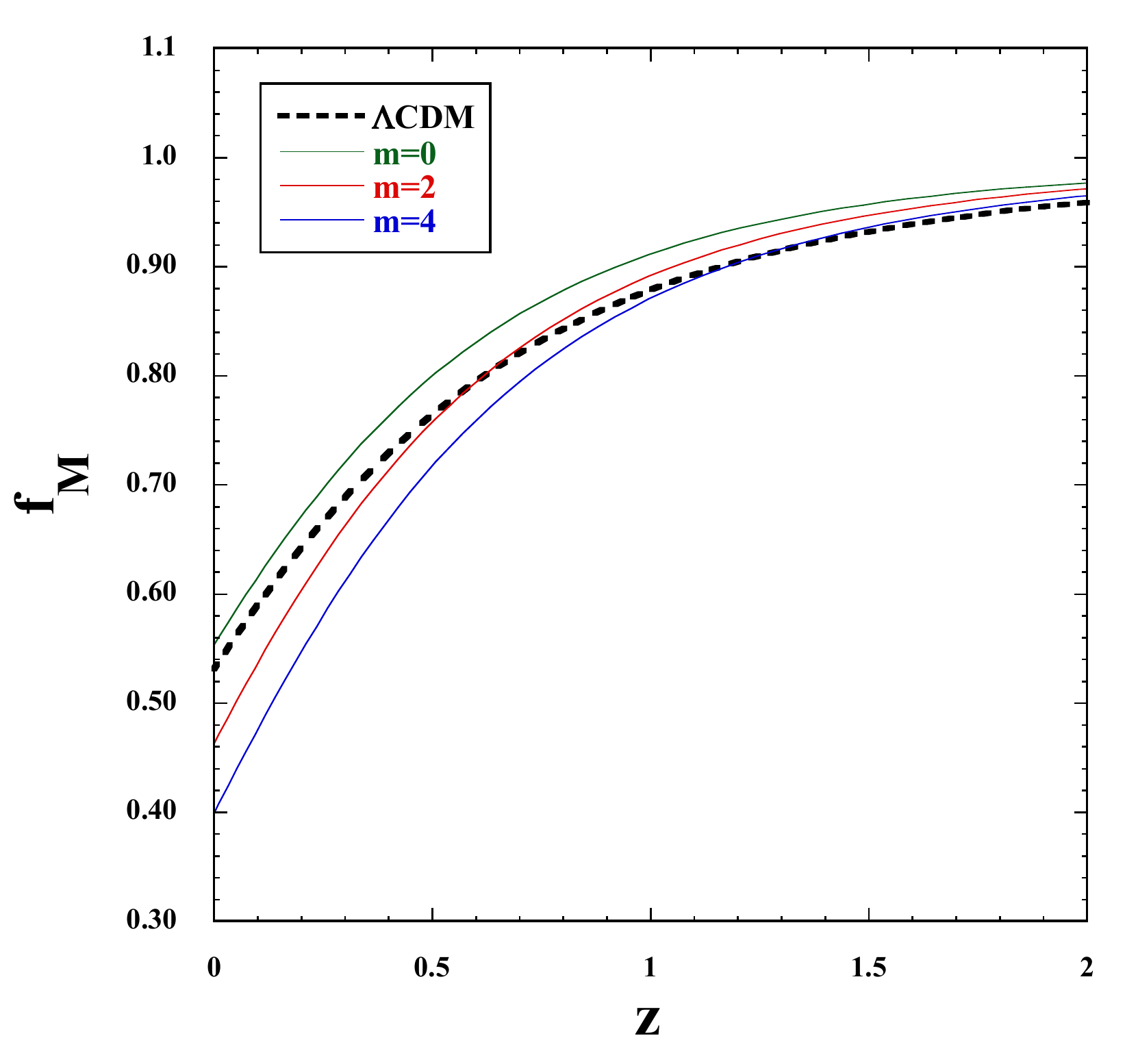}
\end{center}
\caption{Evolution of $f_M=\dot{\delta}_M/(H \delta_M)$ 
versus $z$ for the same background initial conditions 
of density parameters as those used in Fig.~\ref{fig1}. 
The model parameters are 
$s=1/5$, $p_2=1$, and $r_{\beta}=0.05$ with three 
different values of $m$. The dotted line corresponds to 
the evolution of $f_M$ in the $\Lambda$CDM model.
\label{fig3}}
\end{figure}

In the right panel of Fig.~\ref{fig2}, we plot the evolution of $\delta_c$, $\delta_b$, 
$\delta_M$, and $\Phi$ for the same model parameters and background 
initial conditions as those used in the left. 
Here, $\delta_M$ is the total density contrast defined by 
\be
\delta_M=\frac{\Omega_c}{\Omega_c+\Omega_b} \delta_c
+\frac{\Omega_b}{\Omega_c+\Omega_b} \delta_b\,.
\ee
We numerically solve Eqs.~(\ref{deltaceqf}) and (\ref{deltabeqf}) with 
Eqs.~(\ref{Gcc}) and (\ref{Gbb}) derived under the quasi-static 
approximation for linear perturbations deep inside the sound horizon. 
We start to integrate the perturbation equations around the redshift $z=50$ 
by choosing the initial conditions $\delta_c=\delta_c'=\delta_i$ and 
$\delta_b=\delta_b'=\delta_i$. The initial amplitude $\delta_i$ is determined 
by reproducing today's observed matter density contrast 
$\delta_M (z=0)$, where we adopt the Planck2018 best-fit value 
$\delta_M (z=0)=0.811$ \cite{Aghanim:2018eyx}.

Since neither $G_{cc}$ nor $G_{bb}$ depends on the wavenumber $k$, 
the CDM and baryon perturbations exhibit scale-independent growth. 
In Fig.~\ref{fig2}, we observe that the growth of $\delta_c$ is suppressed 
relative to that of $\delta_b$ for the redshift $z \lesssim 1$. 
This behavior is attributed to the gravitational interaction of CDM weaker 
than that of baryons. Since the CDM density is about 
five times as large as the baryon density, the total density contrast 
$\delta_M$ is mostly affected by CDM perturbations 
and hence its growth is suppressed 
in comparison to the standard case with $G_{cc}=G_{bb}=G$. 
This should allow the possibility for alleviating the tension of $\sigma_8$ 
between CDM and low-redshift measurements.

In our theory there is no anisotropic stress, so the gravitational potential 
$\Psi$ and the weak lensing potential $\psi_{\rm WL}=(\Psi-\Phi)/2$
are equivalent to each other, i.e., $\Psi=\psi_{\rm WL}=-\Phi$.
In some models like cubic-order uncoupled scalar Galileons where both $G_{cc}$ and $G_{bb}$ 
are larger than $G$, $|\psi_{\rm WL}|$ grows even after the 
onset of cosmic acceleration \cite{Kobayashi:2009wr,Kimura:2011td}. 
This typically induces a negative ISW-galaxy 
cross-correlation, which is disfavored observationally \cite{Renk:2017rzu}. 
In our coupled GP theory, $G_{cc}$ can be smaller than $G$ at low redshifts, 
so it is possible to avoid the enhancement of $|\psi_{\rm WL}|$. 
In the numerical simulation of Fig.~\ref{fig2}, we observe that 
$\Phi~(=-\psi_{\rm WL})$ decreases at low redshifts.

In Fig.~\ref{fig3}, we show the evolution of the matter growth rate 
$f_M=\dot{\delta}_M/(H \delta_M)$ for three different values of $m$, 
with the other model parameters and initial conditions 
same as those used in Fig.~\ref{fig2}. 
When $m=0$, we have $q_c=1$, $\epsilon_c=0$, and $c_S^2=\hat{c}_S^2$ 
in Eqs.~(\ref{deltaceqf}) and (\ref{Gcc}), so the equation of CDM density 
contrast reduces to the same form as that of baryons with the gravitational 
coupling $G_{cc}=(1+\alpha_{\rm B}^2/\hat{\nu}_S)G$. 
Since $G_{cc}=G_{bb}>G$ in this case,  
the growth rate $f_M$ is larger than that in the $\Lambda$CDM model, 
see Fig.~\ref{fig3}. In contrast, for $m \beta>0$, the CDM gravitational coupling 
$G_{cc}$ can be smaller than $G$ at low redshifts. 
In the numerical simulation of Fig.~\ref{fig3}, the growth rate $f_M$ 
for $m=2$ becomes smaller than that in the $\Lambda$CDM model 
at the redshift $z<0.62$. For increasing $m$, the suppression of $f_M$ 
tends to be more significant, see the case $m=4$ 
in Fig.~\ref{fig3}. Thus, our coupled dark energy model with the 
momentum transfer offers a versatile possibility for realizing the 
weak cosmic growth rate.
When our model is confronted  with the observations
of redshift-space distortions, however, we need to caution that 
the growth rates of $\delta_c$ and $\delta_b$ are different from each other. 
The analysis of how to constrain the model with the redshift-space distortion 
data is left for future work.

\section{Conclusions}
\label{sec6}

We studied the cosmology in coupled cubic-order GP theories given by the action 
(\ref{action}) for the purpose of realizing the weak gravitational interaction 
on scales relevant to the growth of large-scale structures.
The new interaction between the CDM four velocity $u_c^{\mu}$ and 
the vector field $A_{\mu}$, which is weighed by the scalar product 
$Z=-u_c^{\mu}A_{\mu}$, exhibits very different properties in 
comparison to the standard coupled dark energy with 
the energy transfer. The perfect fluids of CDM can be described  
by the Schutz-Sorkin action (\ref{Schutz}), which contains a 
vector density field $J_{c}^{\mu}$ related to the four velocity as
$J_{c}^{\mu}=n_c \sqrt{-g}\,u_c^{\mu}$. 
After deriving general covariant equations of motion in the forms 
(\ref{Ein}) and (\ref{Amueq}), we applied them to the flat FLRW 
background (\ref{metric}).
As we observe in Eqs.~(\ref{conFLRW}) and (\ref{rhoDEeq}), 
the $Z$ dependence in the coupling $f$ does not give rise to 
explicit interacting terms on the right-hand-sides of background 
continuity equations, by reflecting the fact that the interaction 
corresponds to the momentum transfer. 

In Sec.~\ref{sec3}, we derived the second-order actions of tensor, vector, 
and scalar perturbations by choosing the flat gauge given by the 
line element (\ref{permet}). Tensor perturbations propagate in the 
same way as in the standard general relativity, so the theory is 
consistent with the observational bound of speed of gravity 
constrained by the GW170817 event.  
The new interaction does not affect small-scale stability conditions 
of vector perturbations either. 
For scalar perturbations, we obtained the full linear perturbation 
equations of motion and eliminated nondynamical variables 
from the second-order action. The resulting action for dynamical 
perturbations can be expressed in the form (\ref{SS2dynam}), 
which was exploited for the derivation of small-scale stability conditions. 
Under the conditions (\ref{nog1}), (\ref{nog2}), and (\ref{cscon}) 
there are neither ghosts nor Laplacian instabilities, with the 
vanishing effective CDM sound speed. 

In Sec.~\ref{sec4}, we studied the effective gravitational couplings 
for CDM and baryon density perturbations by employing the 
quasi-static approximation for the modes deep inside 
the sound horizon. In our theory, there is no anisotropic stress between 
the two gravitational potentials $\Psi$ and $\Phi$, but 
the $Z$ dependence in $f$ induces the time derivative 
$\dot{\delta}_c$ to $\Phi$ and the longitudinal scalar $\psi$ 
of $A_{\mu}$, see Eqs.~(\ref{PSI}) and (\ref{psi}). 
Differentiating $\Phi$ and $\psi$ with respect to $t$ gives rise 
to the second derivative $\ddot{\delta}_c$ in Eq.~(\ref{delc_evo}) 
of the CDM density contrast. After closing the second-order 
differential equation of $\delta_c$, 
the gravitational coupling for CDM is given by the form (\ref{Gcc}).
In contrast to the baryon gravitational coupling (\ref{Gbb}), 
there are extra terms proportional to $\alpha_{\rm B}$ in $G_{cc}$, 
besides the overall factor $\hat{c}_S^2/(q_S c_S^2)$. 
The terms proportional to $\alpha_{\rm B}$, which correspond to the mixture 
of couplings $G_3(X)$ and $f(Z)$, lead to a value of $G_{cc}$ very different  
from $G_{bb}$ on the de Sitter background, see Eq.~(\ref{GccdS0}).

In Sec.~\ref{sec5}, we proposed a concrete coupled dark energy model 
given by the functions (\ref{model}). 
For the powers (\ref{power}), the background cosmology satisfying the 
relation $\phi^p H={\rm constant}$ ($p>0$) can be realized, with the new 
coupling constant $\beta$  being absorbed into 
the definition of $\Omega_{\rm DE}$. 
In other words, the interaction associated with the momentum transfer 
does not modify the cosmological background of uncoupled GP theories. 
We also showed that the ghosts are absent under the conditions 
(\ref{scon}) and (\ref{mO}). The scalar propagation speed squared 
in each cosmological epoch is given by Eqs.~(\ref{csrad}), (\ref{csma}), 
and (\ref{csds}), which are required to be all positive. The case shown in Fig.~\ref{fig1} 
is an example of the viable cosmology satisfying all the stability conditions.

During the matter dominance, the CDM gravitational coupling $G_{cc}$ is 
expanded in the form (\ref{Gccea}), which can be used to estimate 
the moment after which $G_{cc}$ gets smaller than $G$. 
Provided that the condition (\ref{Omeccon}) is satisfied, the transition to 
the regime $G_{cc}<G$ occurs by today. On the future de Sitter attractor, 
$G_{cc}$ is given by Eq.~(\ref{Gccdsf}), which is always negative in the 
allowed parameter space constrained by the no-ghost and no-strong-coupling 
conditions ($0<p_2\leq1$).
In the numerical simulation of Fig.~\ref{fig2}, which 
corresponds to the power $p_2=1$, $G_{cc}$ enters the region $G_{cc}<G$ 
around $z<1$ and finally approaches the value 
$(G_{cc})_{\rm dS}=-u_{\rm dS}^2 G=-0.225G$. 
In contrast, $G_{bb}$ is always larger than $G$. 
The weak gravitational interaction for CDM leads to 
the suppressed growth of total matter density contrast $\delta_M$, 
see Fig.~\ref{fig2}. The lensing gravitational potential $\psi_{\rm WL}~(=-\Phi)$ 
does not exhibit the enhancement at low redshifts, whose property should be 
consistent with the observations of ISW-galaxy cross-correlations. 
For increasing values of $m$ and $\beta$, 
the growth rates of $\delta_c$ and $\delta_M$ tend to be smaller 
in comparison to the $\Lambda$CDM model, see Fig.~\ref{fig3}.

We thus showed that the coupled GP theories with the momentum transfer 
offers a novel possibility for achieving the weak cosmic growth for CDM, 
in spite of the enhancement of baryon gravitational coupling.
It will be of interest to investigate further whether the interacting model 
proposed in this paper reduces the observational tensions of 
$\sigma_8$ and $H_0$ present in the $\Lambda$CDM model.

\section*{Acknowledgements}

We thank Ryotaro Kase for useful discussions.
ADF thanks Tsujikawa san
laboratory for the warm hospitality at Tokyo University of Science
where this work has started. The work of ADF was supported by
Japan Society for the Promotion of Science Grants-in-Aid for
Scientific Research No.~20K03969. 
ST is supported by the Grant-in-Aid for Scientific Research Fund of the JSPS No.\,19K03854 
and MEXT KAKENHI Grant-in-Aid for Scientific Research on Innovative Areas
``Cosmic Acceleration'' (No.\,15H05890).


\end{document}